\documentclass[%
reprint,
superscriptaddress,
showpacs,preprintnumbers,
 amsmath,amssymb,
prc,
]{revtex4-1}

\usepackage{slashbox}
\usepackage{graphicx}
\usepackage{dcolumn}
\usepackage{bm}

\usepackage{references}
\usepackage{hyperref} 

\usepackage{equationarray}
\usepackage[normalem]{ulem}
\usepackage[usenames]{xcolor}
\usepackage{booktabs}

\newcommand{\be}{\begin{equation}}
\newcommand{\ee}{\end{equation}}
\newcommand{\ba}{\begin{eqnarray}}
\newcommand{\ea}{\end{eqnarray}}
\newcommand{\bd}{\begin{displaymath}}
\newcommand{\ed}{\end{displaymath}}

\usepackage{ulem} 

\begin{document}


\title{Enforcing causality in nonrelativistic equations of state at finite temperature}

\author{Constantinos Constantinou}
\email{c.constantinou@fz-juelich.de }
\affiliation{Institute for Advanced Simulation, Institut f\"{u}r Kernphysik, and J\"{u}lich Center \\
for Hadron Physics, Forschungszentrum J\"{u}lich, D-52425 J\"{u}lich, Germany}

\author{Madappa Prakash}
\email{prakash@ohio.edu}
\affiliation{Department of Physics and Astronomy, Ohio University, Athens, OH 45701}

\date{\today}

\begin{abstract}

We present a thermodynamically consistent method by which equations of state based on nonrelativistic potential models can be 
modified so that they respect causality at  high densities, both at zero and finite temperature (entropy). We illustrate the
application of the method using  the high density phase  parametrization of the well known APR model in its pure neutron
matter  configuration as an example. We also show that, for models with only contact interactions, the adiabatic
speed of sound is independent of the temperature  in the limit of very large temperature. This feature is
 approximately valid for models with finite-range interactions as well, insofar as the temperature dependence they
 introduce to the Landau effective mass is weak. In addition, our study  
 reveals that in first principle nonrelativistic models of hot and dense matter, contributions from
   higher than two-body interactions must be screened at high density to preserve causality. 
\\

\end{abstract}

\maketitle


\section{Introduction}

The precise determination of neutron star masses close to 2$M_\odot$ \cite{Demorest10,Antoniadis13}, prospects of observing gravitational waves (GW's)
from mergers involving binary neutron stars as in the recent  detection of GW's from mergers of binary black holes \cite{Abbott16a,Abbott16b}, and the
hope of observing a nearby core-collapse supernova (SN) with the several neutrino observatories currently in place have greatly strengthened the study
of dense matter physics.  Central to this study is the equation of state (EOS) of dense matter at both zero and finite temperature. Depending on the
values of the baryon densities, $n$, and temperatures, $T$, reached in core-collapse supernovae, neutron stars from their birth to old age, and  mergers
of compact binary stars, several phases of matter may be encountered.  At high densities and/or temperatures, these phases may consist of strangeness-bearing
hadrons and/or quark matter \cite{Prakash97}.  

Large-scale computer simulations of the astrophysical phenomena mentioned above employing the microscopic physics input of model EOS's from both
nonrelativistic and relativistic approaches have indicated the ranges of $n/n_s$, where $n_s\simeq0.16~{\rm fm}^{-3}$ is the nuclear saturation density,
$T$, and the net electron fraction $Y_e=n_e/n$  encountered.  To enable simulations,  EOS's that range over $n/n_s$ up to 10, $T$ up to 200 MeV, and $Y_e$
up to 0.6 are required.  These conditions imply an entropy per baryon $S$ (in units of Boltzmann's constant $k_B$) of up to 200. In varying amounts the
entropy is shared between the hadrons, leptons, and photons. For EOS's with only nucleonic components, $S_{nuc}$ of up to 4-5 is not uncommon.    
In the homogeneous phase ($n \ge 0.1$ fm$^{-3}$) and using the EOS of APR \cite{APRus}, $S \le 30$ with nucleons contributing about 5, leptons 15, and
photons 10. The highest density at which $S \sim 200$ is about 0.01 fm$^{-3}$ (at $T$=200 MeV) with $95\%$ of the contributions coming from leptons and
photons. The dependence of these numbers on the charge/lepton fraction is, generally, very weak.   

The focus of this paper is on the adiabatic speed of sound $c_s$ in matter at high density and temperature.  In hydrodynamical simulations, $c_s$ represents
a physical scale which controls the macroscopic evolution of matter. Thus, a quantitative knowledge of how $c_s$ varies with $n$ and $T$ (or $S$) in models
of hot, dense matter can shed light on the time development of involved hydrodynamical simulations. In physical systems, $c_s$ cannot exceed the speed of light $c$.
Nevertheless, many of the EOS's used to describe nucleonic matter have nonrelativistic underpinnings and therefore do not conform to the 
requirement of causality. (Relativistic field-theoretical approaches to dense matter  inherently respect causality, and will not be addressed further in this work.)  
For some nonrelativistic EOS's, typically the softer ones, which struggle to support neutron star (NS) masses $\sim 2 M_{\odot}$, the causality requirement is
only violated for densities beyond their scope. The stiffer EOSs however, can become problematic for densities and temperatures for which hadronic matter is
expected to persist within a star. 

Repulsive contributions to the energy per particle $E(u=n/n_s)$ that vary faster than linear in $u$ give rise to acausal behavior at high densities. Thus,
higher than two-body forces found necessary to achieve saturation at the empirical $n_s$ with the empirical binding energy of symmetric nuclear matter (SNM)
must screen themselves with progressively increasing density to ensure causal behavior.  Following the suggestion in Ref. \cite{Bludman80}, causality was
maintained through the use of $Bu^\sigma/(1+B^\prime u^{\sigma-1})$, where $B$ is a constant of dimension energy and $B^\prime$ is a dimensionless constant appropriately
chosen to have $c_s$ approach $c$ from below, in the explorative study of Ref. \cite{Prakash88}.  This implied self-screening of repulsive interactions,
while desirable, is not always guaranteed in nonrelativistic  potential-model calculations. This issue is particularly relevant to modern microscopic calculations
of the EOS of SNM and pure neutron matter (PNM), such as the quantum Monte Carlo  \cite{Armani:2011mn,Gandolfi:2011xu,Wlazlowski:2014jna} and chiral effective field theory 
approaches \cite{Gezerlis10,Hebeler10,Coraggio:2012ca,Tews13,Togashi13,Lynn16,Epelbaum:2008vj,Hu:2016nkw} in which the role of three-body forces at $T=0$ have been examined. Owing to
inherent technical difficulties, calculations have been limited up to about 2$n_s$  in both of these approaches. To calculate the structural properties of NS's,
the EOS's have been extrapolated beyond $\sim 2n_s$ through the use  of piece-wise polytropes that respect causality (thus screening the influence of 3-body forces
present at $n<2n_s$, possibly prematurely for higher densities). This polytropic extrapolation, while satisfactory at $T=0$ on a practical level, cannot however be
extended to finite temperature unless the effects of temperature on the EOS are known a priori. 

Here, we present a thermodynamically consistent method to maintain causality for EOS's that become acausal at both zero and finite temperature. While such a
method is available in the literature for zero temperature \cite{Nauenberg73,LPMY90,APRus}, a method to encompass the influence of temperature on $c_s$ has
not received much attention (the method presented in Appendix E of our earlier work in Ref. \cite{APRus} contained an inadvertent error, which is corrected in
this work).   We illustrate the application of the method using  a few chosen models \cite{LS,sly4,APR} that become acausal at  high density and temperature.
These models have distinctly different behaviors in their nucleon effective masses as functions of  density. For simplicity, results for PNM are shown in all
cases with the generalization to a multi-component system indicated in the text. We stress, however, that the applicability of the method proposed is not limited
to the class of models chosen for illustration. As long as the relevant thermodynamic variables such as the energy, pressure, and chemical potential for any
model are available for all densities and temperatures of interest, the method can be used to render the EOS causal and to satisfy the thermodynamic identity.   

The paper is organized as follows. In Sec. \ref{General},  the adiabatic speed of sound $c_s$ is defined in terms of thermodynamic quantities  characteristic
of an EOS  at both zero and  finite temperature. This section also contains a discussion of the behavior of $c_s$ in the limiting cases of degenerate and
nondegenerate bulk matter. The method devised  to implement causality for EOS's that become acausal at high densities  and temperatures is described in
Sec. \ref{Implementation}.  In Sec. \ref{Numerical}, the numerical procedure to enforce causality for models with contact interactions is detailed. Results in
the case of PNM for these models are presented in Sec. \ref{Results}  both at zero and finite temperature. Additionally, results for a model in which contributions
from higher than two-body  interactions are screened to prevent an acausal behavior are presented. Section \ref{Conclusions} presents a summary and conclusions.

\section{General Considerations}
\label{General}

For small-amplitude perturbations, the velocity $v$ of fluid particles obeys the wave equation \cite{LL-fluids}
\begin{equation}
  \frac{\partial^2v}{\partial t^2}-c_s^2~\frac{\partial^2v}{\partial x^2} = 0
\end{equation}
whose solution $f(x-c_st)$ represents longitudinal sound wave propagation with speed $c_s$ under the condition of adiabatic motion for which
${\partial S}/{\partial t} = 0$, where $S$ is the entropy. Thus small density fluctuations in a compressible fluid
propagate at the speed of sound given by \cite{Weinberg71,Guichelaar74,Prak-Notes}
\begin{eqnarray}
  \left(\frac{c_s}{c}\right)^2 &=& \left.\frac{\partial P}{\partial \epsilon}\right|_S
                               = \frac{\left.\partial P/\partial n\right|_S}{\left.\partial \epsilon/\partial n\right|_S} \\
                               &=& \frac{1}{\mu+m}\left.\frac{\partial P}{\partial n}\right|_S
                               = \frac{K_s}{9(\mu+m)} = \frac{\Gamma_S P}{h + mn} \,,
\end{eqnarray}
where $P$ is the pressure, $\epsilon$ is the energy density, $n$ is the number density, $h=nE+P$ is the specific enthalpy density, $\mu$ is the chemical
potential, $K_S = 9 \left.\partial P/\partial n\right|_S$ the adiabatic incompressibility, and $\Gamma_S = \left.\partial \ln P/\partial \ln n\right|_S$
the adiabatic index. In the variables $(n,T)$, we can also express $c_s$ as \cite{Guichelaar74,MDYI}
\begin{equation}
  \left(\frac{c_s}{c}\right)^2 =  \frac{C_P}{C_V}\frac{n}{h+mn}\left.\frac{\partial P}{\partial n}\right|_T \,,
\end{equation}
where $C_P$ and $C_V$ are the specific heats at constant pressure and volume, respectively. 

We will begin our analysis with nonrelativistic models that are described by the generic Hamiltonican density 
\begin{equation}
    \mathcal{H} = \frac{\hbar^2}{2}\frac{\tau(n,S)}{m^*(n)} + V(n) \,,  
    \label {Hdensity}
\end{equation}
where the first term is the kinetic energy density and the second is the potential energy density. The quantity $m^*$ is the Landau effective mass defined
at the Fermi surface by $m^*(n) = p_F~(\partial \epsilon_p/\partial p)|_{p_F}$, where $\epsilon_p$ is the single particle spectrum and $p_F$ is the Fermi momentum.
Models that employ contact interactions such as Skyrme models, the APR model, and other microscopic models that employ the effective mass approximation as
indicated above are examples of the representation in Eq. (\ref{Hdensity}). As our discussion proceeds, we will  consider  other cases, {\em e.g.}, models in
which finite-range interactions at various levels of sophistication are considered and in which additional complications are encountered. 
Some physical insight is gained by examining the behavior of $c_s$ in the limiting situations of degenerate and nondegenerate  bulk matter to which we turn below. 

\subsection*{Degenerate case}

For conditions such that $T/T_F\ll1 $ (or, equivalently $S\leq 1$), where $T_F=p_F^2/(2m^*)$  is the Fermi temperature in nonrelativistic models, degenerate conditions prevail.   
In this case, the leading-order Fermi liquid theory (FLT) expressions for  the thermal components of the pressure and energy density are given by \cite{flt,Prakash87}
\begin{eqnarray}
P_{th}(n,T) &=& \frac 23 naT^2Q \quad {\rm and} \quad  \epsilon_{th} (n,T) = naT^2  \\ 
{\rm with} \quad Q &=& 1-\frac{3}{2}\frac{n}{m^*}\frac{dm^*}{dn}  \,,
\label{Q}
\end{eqnarray}
where $a=\pi^2/(4T_F)$ is the level density parameter.  Utilizing the leading-order FLT result $S=2aT$, we get
\begin{eqnarray}
P_{th}(n,S) &=& \frac {S^2}{6} \frac {nQ}{a} = \frac {2S^2}{3\pi^2}~  nT_FQ \quad {\rm and} \\ 
\quad \epsilon_{th}(n,S) &=& \frac {S^2}{4} \frac na = \frac {S^2}{\pi^2}~ nT_F\,, 
\end{eqnarray}
from which the density derivatives at constant $S$
\begin{eqnarray}
\left.\frac {dP_{th}}{dn}\right|_S &=& \frac {P_{th}}{n} \left(1 + \frac 23 Q + \frac nQ \frac {dQ}{dn}\right) \,, \nonumber \\
\left.\frac {d\epsilon_{th}}{dn}\right|_S &=& \frac {\epsilon_{th}}{n} \left(1 + \frac 23 Q\right) \quad {\rm and} \,, \nonumber \\
\left.\frac {dT}{dn}\right|_S &=& \frac 23 \frac {TQ}{n} 
\label{thderivs}
\end{eqnarray}
are easily obtained. Putting together the other components in the total $P$ and $\epsilon$, we arrive at
\begin{eqnarray}
\left. \frac {dP}{dn}\right|_S &=& \frac 25 T_F\left( 1+ \frac 23 Q \right) + n \frac {d^2V}{dn^2} + \left.\frac {dP_{th}}{dn}\right|_S  \nonumber \\
\left. \frac {d\epsilon}{dn}\right|_S &=& m + \frac 35 T_F \left( 1+ \frac 23 Q \right) + \frac {dV}{dn} 
+  \left.\frac {d\epsilon_{th}}{dn}\right|_S \,,
\label{totderivs}
\end{eqnarray}
where the first term in the first equation above and the second term in the second represent the $T=0$ results of the kinetic parts.
These results allow us to appreciate how $c_s^2$ is governed by physical quantities in special circumstances. \\

\noindent (i) The case when $V=0$: When the thermal contributions can be regarded as small ($S\leq 1$) compared to their zero temperature counterparts,
\begin{eqnarray}
 c_s^2 &\simeq&  \frac {2}{5} \frac {T_F}{m} \left( 1+ \frac 23 Q \right) \\
 &=& \frac 15 \left(\frac{p_F}{m^*}\right)^2 \frac {m^*}{m} \left( 1+ \frac 23 Q \right) \,. 
 \label{special1}
\end{eqnarray}
The physical scale of $c_s$ here is the velocity at the Fermi surface $v_F = p_F/m^*$ modified by $m^*/m$ and its logarithmic derivative with respect to density
contained in $Q$. For $m^*(n)=m$, the ideal gas value of $c_s^2=(1/3) (p_F/m)^2$ is recovered from Eq. (\ref{special1}).\\

\noindent (ii) Density-dependent $V(n)$:  In order to achieve equilibrium at the empirical nuclear  density $n_0\simeq0.16~{\rm fm}^{-3}$ with the empirical energy per
particle of $\simeq -16$ MeV, and to support the precisely determined neutron star masses of 2$M_\odot$'s, models of dense nuclear matter have employed contributions
from beyond 2-body forces in $V(n)$ that vary as $n^{2+\epsilon}$. If these contributions persist at  densities $n\gg n_0$,  and dominate over the other contributions
including the thermal parts ($S\leq 1$), causality is bound to be violated. Consider, for example, $V(n) \propto n^\sigma$ for which 
\begin{eqnarray}
\left(\frac{c_s}{c}\right)^2 \simeq \frac {n~d^2V/dn^2}{dV/dn} = \sigma -1 \,, 
\label{special2}
\end{eqnarray}
which for $\sigma \geq 2$ renders $(c_s/c)^2 \geq 1$. \\

\noindent (iii) Additional contributions to the density-dependent $V(n)$: As apparent from Eq. (\ref{totderivs}), both the $S=0$ and $S\neq 0$ terms contribute in determining
the magnitude of $(c_s/c)^2$. The interplay between  these terms is also determined by  $m^*(n)$ and its density derivatives as well as by $Q(n)$ and its derivatives.
In Skyrme-like models in which  $m^*/m=(1+\beta n)^{-1}$ with $\beta$ a constant, the kinetic energy density, $\varepsilon_{kin}\propto n^{5/3}(1+\beta n)$, so that at some high
$n$ the $n^{8/3}$ term dominates causing the EOS to become acausal. In some cases, acausality can set in at lower densities for $S\neq 0$ than for $S=0$.  
A quantitative discussion of results from models with different behaviors of $m^*/m~{\rm vs}~n$ will  be deferred to  Sec. \ref{Results}.

\subsection*{Nondegenerate case}

At zero temperature (entropy), $c_s$ is a function of just the density $n$. As we show below, the same is true for nonrelativistic models with only contact interactions
in the limit of very large entropy/temperature, {\it i.e.,}  the extreme nondegenerate limit. Here, the entropy is given by the Sackur-Tetrode relation
\begin{equation}
  S = \frac{5}{2}-\ln\left[\left(\frac{2\pi\hbar^2}{m^*T}\right)^{3/2}\frac{n}{2}\right] \,,
   \label{snd}
\end{equation}
where $m^*$ is the density-dependent Landau effective mass. Solving Eq.~(\ref{snd}) for the temperature, we get
\begin{eqnarray}
  T = \frac{2\pi\hbar^2}{m^*}\left(\frac{n}{2}\right)^{2/3}\exp\left(\frac{2}{3}S-\frac{5}{3}\right)\,, ~~
    \left.\frac{\partial T}{\partial n}\right|_S = \frac{2}{3}\frac{TQ}{n}\,, \nonumber \\
\end{eqnarray}
with $Q$ given by Eq. (\ref{Q}). In this case, thermal effects dominate over cold matter contributions (exclusive of rest-mass) to thermodynamic properties. Consequently, 
\begin{equation}
  \left(\frac{c_s}{c}\right)^2 \stackrel{S\gg1}{\longrightarrow} \frac{\left.\partial P_{th}/\partial n\right|_S}
                                                                        {m + \left.\partial \varepsilon_{th}/\partial n\right|_S} \,.
\label{csnondeg}                                                                        
\end{equation}
For nonrelativistic contact-interaction models in the nondegenerate limit, 
\begin{eqnarray}
  P_{th} &=& nTQ\,,  \quad \left. \frac {\partial P_{th}}{dn}\right|_S  = TQ\left( 1 + \frac 23 Q + \frac nQ \frac {dQ}{dn}\right)  \\
   \varepsilon_{th} &=& \frac{3}{2}nT\,, \quad \left. \frac {\partial \varepsilon_{th}}{dn}\right|_S = \frac 32 T \left(1+\frac 23 Q \right) \,.
\end{eqnarray}
When the mass term dominates over the thermal part in the denominator of Eq. (\ref{csnondeg}), and $Q\simeq 1$ as is the case when the effects of interactions are small,   
\begin{equation}
  \left(\frac{c_s}{c}\right)^2 \stackrel{S\gg1}{\longrightarrow} ~\frac 53 ~\frac Tm 
\end{equation}
which is the result for one-component classical gases. The physical scale of $c_s$ here is the thermal velocity of particles.   
In the case that the thermal component of  the denominator in Eq. (\ref{csnondeg}) dominates over the mass, 
\begin{equation}
  \left(\frac{c_s}{c}\right)^2 \stackrel{S\gg1}{\longrightarrow} ~\frac{2}{3} ~\frac{Q\left(1+\frac{2}{3}Q+\frac nQ\frac{dQ}{dn}\right)}{1+\frac{2}{3}Q} \,,  \label{csindept}
\end{equation}
that is, the temperature/entropy dependence drops out with the result $(c_s/c)^2 \simeq 2/3$ for $Q\simeq 1$.  
It must be emphasized that this result is obtained only at very high temperatures for which the use of nonrelativistic considerations becomes questionable.  

\subsection*{Models with finite-range forces}

Finite-range forces introduce momentum dependences (other than $p^2$) to the single-particle potential which in turn cause it to acquire a
temperature dependence \cite{MDYI}. The effects of these interactions can still be collected in a density- and temperature-dependent function $m^*(n,T)$
which, however, can no longer be identified with the Landau effective mass. Nevertheless, if this $T$-dependence is weak, then
\begin{eqnarray}
  m^*(n,T) &\simeq& m^*(n,0) + T \left.\frac{\partial m^*(n,T)}{\partial T}\right|_{T=0} + \ldots  \\
          &\equiv& m^*(1+bT)    \label{mlimit}
\end{eqnarray}
where $m^* = m^*(n,0)$ is the Landau mass and $b(n)\equiv (1/m^*) \left.\partial m^*(n,T)/\partial T\right|_{T=0}$ such that $bT\ll1$.
Combining Eq. (\ref{snd}) with $m^*\rightarrow m^*(n,T)$ and Eq. (\ref{mlimit}), and expanding in a Taylor series for $bT\ll 1$, we find
\begin{equation}
  S \simeq \frac{5}{2}-\ln\left[\left(\frac{2\pi\hbar^2}{m^*T}\right)^{3/2}\frac{n}{2}\right] + \frac{3bT}{2} \label{snd1} \,.
\end{equation}
Perturbative inversion of Eq. (\ref{snd1}) yields, in the second recursion,
\begin{eqnarray}
  T &\simeq& \frac{2\pi\hbar^2}{m^*}\left(\frac{n}{2}\right)^{2/3}\exp\left[\frac{2}{3}S-\frac{5}{3}\right. \nonumber \\
    &-& \left.\frac{2\pi\hbar^2}{m^*}b\left(\frac{n}{2}\right)^{2/3}\exp\left(\frac{2}{3}S-\frac{5}{3}\right)\right]  \label{T1}\,.
\end{eqnarray}
We now substitute $2\pi\hbar^2(n/2)^{2/3}\exp(2S/3-5/3)$ by $m^*(n,T)$, then replace  $m^*(n,T)$ by
$m^*(1+bT)$ to get 
\begin{eqnarray}    
T    &\simeq& \frac{2\pi\hbar^2}{m^*}\left(\frac{n}{2}\right)^{2/3}\exp\left[\frac{2}{3}S-\frac{5}{3}-\frac{bTm^*(n,T)}{m^*}\right] \label{T2} \nonumber \\
    &\simeq& \frac{2\pi\hbar^2}{m^*}\left(\frac{n}{2}\right)^{2/3}\exp\left[\frac{2}{3}S-\frac{5}{3}-bT(1+bT)\right] \label{T3} \,.
    \end{eqnarray}
Now dropping $b^2T^2$ and expanding the exponential for small $bT$,
\begin{equation}    
T  \stackrel{\stackrel{bT\ll1}{S\gg1}}{\longrightarrow} \frac{2\pi\hbar^2}{m^*}\left(\frac{n}{2}\right)^{2/3}\exp\left(\frac{2}{3}S-\frac{5}{3}\right) \,. \label{T5}
\end{equation}  
This result shows that in the extreme nondegenerate limit, finite-range force models with weak $T$-dependence in their $m^*$'s will behave similarly to zero-range models
and thus they will also obey Eq. (\ref{csindept}).

\section{Implementation of Causality}
\label{Implementation}

The general approach described below is more conveniently applied in the variables $(n,S)$ that are natural to the speed of sound as opposed to $(n,T)$ commonly used
in tabulations of EOS properties. Working with the former set allows us to carry out all calculations analytically circumventing the need for numerical integration.

Causality is preserved as long as the speed of sound $c_s$ is less than or equal to the speed of light $c$:
\begin{equation}
\left(\frac{c_s}{c}\right)^2 \equiv \beta = \left.\frac{\partial P}{\partial \epsilon}\right|_S
                     = \left.\frac{\partial P}{\partial n}\right|_S\left(\left.\frac{\partial \epsilon}{\partial n}\right|_S\right)^{-1}\le 1 .
\label{beta}
\end{equation}
Here the total energy density $\epsilon$ is inclusive of the internal energy density $\varepsilon$ and the rest-mass energy density $mn$:  
\begin{equation}
  \epsilon = \varepsilon + mn \,.
\end{equation}
By making use of
\begin{equation}
  P = n^2\left.\frac{\partial(\epsilon/n)}{\partial n}\right|_{S,N} = n\left.\frac{\partial \varepsilon}{\partial n}\right|_{S,N}-\varepsilon 
  \label{Pres}  
\end{equation}
and
\begin{equation}
    \left.\frac{\partial P}{\partial n}\right|_{S,N} = n \left.\frac{\partial^2\varepsilon}{\partial n^2}\right|_{S,N}  
\end{equation}
where $N$ is the number of nucleons in the system, we write Eq. (\ref{beta}) as a second-order differential equation (DE)
\begin{equation}
  \left.\frac{\partial^2\varepsilon}{\partial n^2}\right|_{S,N} -\frac{\beta}{n}\left.\frac{\partial \varepsilon}{\partial n}\right|_{S,N} = \frac{\beta m}{n}. \label{2DE}
\end{equation}
Thus, in addition to entropy conservation, in our approach we must impose the condition of baryon number conservation and, in the case of multicomponent systems, fixed composition.
Equation (\ref{2DE}), can be reduced to a first-order DE
\begin{equation}
\left.\frac{\partial \xi}{\partial n}\right|_{S,N} - \frac{\beta}{n}\xi = \frac{\beta m}{n}  \label{1DE}  
\end{equation}
by setting 
\begin{equation}
  \xi = \left.\frac{\partial \varepsilon}{\partial n}\right|_{S,N} = \mu + TS \,.   
  \label{xi}
\end{equation}
Note that the combination of Eqs. (\ref{Pres}) and (\ref{xi}) yields the thermodynamic identity $\varepsilon + P = n\mu + Ts$. The solution of Eq. (\ref{1DE})
requires that $\beta$ is mapped to some function $\beta_f(n,S)\le 1 ~\forall (n,S)$. This implies that the causality-fixing density $n_f$ obtained from
\begin{equation}
  \beta(n,S)-\beta_f(n,S) = 0 \label{nfs}
\end{equation}
is an entropy-dependent function. 

The approach of $c_s$ to $c$ depends on the choice of $\beta_f(n,S)$. For our illustrative calculations below, some choices of $\beta_f(n,S)$ are considered. 

\subsection*{Density-independent $\beta_f(n,S)$}

For such a constant $\beta_f$, the integrating factor corresponding to Eq. (\ref{1DE}) is given by
\begin{equation}
f(n) = \exp\left(-\beta_f \int\frac{dn}{n}\right) = n^{-\beta_f} ,
\end{equation}
and has the property
\begin{equation}
\frac{d}{dn}[n^{-\beta_f}\xi] = n^{-\beta_f}~ \frac{\beta_f m}{n}. \label{ifac}    
\end{equation}
Integration of Eq. (\ref{ifac}) leads to
\begin{equation}
\xi = \left.\frac{\partial \varepsilon}{\partial n}\right|_{S,N} = -m+c_1n^{\beta_f} , \label{mu1}   
\end{equation}
where $c_1$ is a constant of integration. A second integration results in
\begin{equation}
\varepsilon = -mn + \frac{c_1n^{\beta_f+1}}{\beta_f+1} + c_2  \label{eden1}
\end{equation}
with another constant of integration $c_2$, and therefore
\begin{equation}
P = c_1\frac{\beta_f}{\beta_f+1}n^{\beta_f+1} - c_2 . \label{pres1}
\end{equation}
The constants $c_1$ and $c_2$ are determined by the boundary conditions
\begin{eqnarray}
\varepsilon[n_f(S),S] &=& \varepsilon_f(S) \label{ef}\\
P[n_f(S),S] &=& P_f(S) ,\label{pf}
\end{eqnarray}
where $n_f$ is the causality fixing density, defined by Eq. (\ref{nfs}). The functional forms of $\varepsilon(n,S)$ and $P(n,S)$ are those obtained 
from the original Hamiltonian density.

From Eqs. (\ref{ef}) and (\ref{pf}),  we get
\begin{eqnarray}
c_1 &=& \frac{\varepsilon_f+mn_f+P_f}{n_a^{\beta_f+1}}   \label {c1}  \\
c_2 &=& \frac{1}{\beta_f+1}[\beta_f(\varepsilon_f+mn_f)-P_f]. \label{c2}
\end{eqnarray}
Thus the energy density and the pressure are given by
\begin{eqnarray}
  \varepsilon &=& -mn + \frac{(\varepsilon_f+mn_f+P_f)}{\beta_f+1}\left(\frac{n}{n_f}\right)^{\beta_f+1} \nonumber \\
  &+& \frac{\beta_f(\varepsilon_f+mn_f)-P_f}{\beta_f+1} \label{veps} \\
  P &=& \frac{\beta_f}{\beta_f+1}(\varepsilon_f+mn_f+P_f)\left(\frac{n}{n_f}\right)^{\beta_f+1} \nonumber \\
  &-& \frac{\beta_f(\varepsilon_f+mn_f)-P_f}{\beta_f+1}. \label{pp}
\end{eqnarray}
The chemical potential $\mu$ is straightforwardly obtained from $\mu=\xi-TS$. Equations (\ref{veps}) and (\ref{pp}) can be used for $n \ge n_f$ with a
fixed $\beta_f\le 1$ so that causality is never violated and such that the thermodynamic identity  is obeyed thus ensuring thermodynamic consistency.
At this stage, we must reiterate the point that $\varepsilon$ and $P$ as given in Eqs. (\ref{veps}) and (\ref{pp}) are functions of $(n,S)$. The switch
to $(n,T)$ is easily achieved by setting
\begin{eqnarray}
  \varepsilon(n,T) &=& \varepsilon[n,S(n,T)]  \\
  P(n,T) &=& P[n,S(n,T)] \\
  \beta(n,T) &=& \beta[n,S(n,T)].
\end{eqnarray}

Note that the procedure outlined above for $T\neq 0$ closely mirrors that for $T=0$ described in Appendix E of our earlier work in Ref. \cite{APRus}, but with the
use of appropriate quantities at finite $T$. The method outlined in  Ref. \cite{APRus} for $T\neq 0$ was flawed in that Eq. (E18) there defining the chemical potential
lacked a term involving $TS$, that is, $\mu$ was taken to be $\partial \varepsilon / \partial n |_{S,N}$ instead of the correct $\partial \varepsilon / \partial n |_{S,V}$
where $n=N/V$. Equation  (\ref{xi}) in this work corrects that error. Moreover, the assumption that $C_P/C_V= {\rm constant}$ was made, which is only true in the
degenerate limit ($S \le 1$) where $C_P/C_V\simeq 1$.  \\

\subsection*{Density-dependent $\beta_f(n,S)$}

We emphasize that $\beta_f$ need not be a constant. Consider, for the purposes of illustration, the function
\begin{equation}
  \beta_f(n,S) = a_1 + \frac{a_2n^{a_3}}{1+a_4n^{a_3}}
\end{equation}
where the $a_i$ are real numbers; $a_1$ and $a_3$ are unitless while $a_2$ and $a_4$ have units fm$^{3a_3}$. For this $\beta_f$ to approach 1 from
below they must all be positive and $a_1+a_2/a_4=1$; $a_1 >0$ also ensures that Eq. (\ref{nfs}) always has a solution. Moreover, if the fraction $a_2/(a_3a_4)$
is an integer, then $\xi$, $\varepsilon$, and $P$ are relatively simple functions of the density.

For example, if we choose $a_1=1/2$, $a_2=2$, $a_3=1$, and $a_4=4$ then
\begin{eqnarray}
 \beta_f &=& \frac{1}{2} + \frac{2n}{1+4n}  \label {beta2}\\    
 \xi &=& -m + c_1n^{1/2}(1+4n)^{1/2} \\
 \varepsilon &=& -mn + \frac{c_1}{16}\left[n^{1/2}(1+4n)^{1/2}(1+8n) \right. \nonumber \\
             &-& \left. \frac{1}{2}\sinh^{-1}(2n^{1/2})\right] + c_2   \\
P &=& - \frac{c_1}{16}\left[n^{1/2}(1+4n)^{1/2}(1-8n) \right. \nonumber \\
  &-& \left. \frac{1}{2}\sinh^{-1}(2n^{1/2})\right] - c_2 \,.
\end{eqnarray}
For the choice $a_1=4/5$, $a_2=2$, $a_3=1/5$, and $a_4=10$, we get
\begin{eqnarray}
  \beta_f &=& \frac{4}{5} + \frac{2n^{1/5}}{1+10n^{1/5}}  \\
  \xi &=& -m + c_1(n^{4/5}+10n)  \\
  \varepsilon &=& -mn + \frac{5}{9}c_1(n^{9/5}+9n^2) + c_2 \\
  P &=& c_1\left(\frac{4}{9}n^{9/5}+5n^2\right) -c_2 .
\end{eqnarray}
Of course, many other possibilities exist for the $a_i$ above, as well as for the generic functional form of $\beta_f$.

\section{Numerical Notes}
\label{Numerical}
Here we describe the procedure to calculate $(c_s/c)^2$ for the Hamiltonian density in Eq. (\ref{Hdensity}) for conditions of arbitrary degeneracy. The analytical results
obtained  in Sec. \ref{General} for the degenerate and nondegenerate cases serve as a check for the results obtained in this section. To calculate the finite-entropy
properties corresponding to  Eq. (\ref{Hdensity}), we employ the Johns, Ellis and Lattimer (JEL)  \cite{jel} scheme in which the Fermi-Dirac  integrals 
\begin{equation}
F_{\alpha} = \int_0^\infty \frac {x^\alpha}{e^{x-\psi}  + 1}\,dx
\end{equation}
are expressed as algebraic functions of a single parameter $f$ related to the entropy via
\begin{equation}
  S = \frac{5}{3}\frac{F_{3/2}(f)}{F_{1/2}(f)} - \psi(f)  \label{sjel}
\end{equation}
where \cite{APRus}
\begin{eqnarray}
  F_{3/2}(f) &=& \frac{3f(1+f)^{1/4-M}}{2\sqrt{2}}\sum_{m=0}^M p_m f^m  \label{f32}\\  
  F_{1/2}(f) &=& \frac{f(1+f)^{1/4-M}}{\sqrt{2(1+f/a)}}\sum_{m=0}^M p_m f^m     \nonumber  \\
  &\times& \left[1+m-\left(M-\frac{1}{4}\right)\frac{f}{1+f}\right]  \label{f12}\\
  \psi(f) &=& \frac{\mu(n,S)-V(n)}{T(n,S)}    \nonumber \\
  &=& 2(1+f/a)^{1/2} + \ln\left[\frac{(1+f/a)^{1/2}-1}{(1+f/a)^{1/2}+1}\right] \,. \label{psif}
\end{eqnarray}
The values of the coefficients appearing in Eqs. (\ref{f32})-(\ref{psif}) are $a=0.433$, $M=3$, $p_0=5.34689$, $p_1=16.8441$, $p_2=17.4708$, and $p_3=6.07364$. 
We note that the $F_{\alpha}$ are connected via their derivatives with respect to $\psi$ according to $\partial F_{\alpha}/\partial \psi =  \alpha F_{(\alpha-1)}$.
The JEL scheme enables a rapid and an accurate evaluation of the thermodynamic quantities preserving  thermodynamic consistency. 

The kinetic energy density $\tau$ and the number density $n$ are related to $F_{3/2}$ and $F_{1/2}$, respectively:
\begin{eqnarray}
  \tau(n,S) &=& \frac{\gamma}{2\pi^2}\left[\frac{2m^*(n)T(n,S)}{\hbar^2}\right]^{5/2}F_{3/2}[f(S)]  \label{tau}\\
          n &=& \frac{\gamma}{2\pi^2}\left[\frac{2m^*(n)T(n,S)}{\hbar^2}\right]^{3/2}F_{1/2}[f(S)] \,, \label{nden}
\end{eqnarray}
where $\gamma=1(2)$ for PNM(SNM) and $f(S)$ is the solution of Eq. (\ref{sjel}). From Eq. (\ref{nden}), it follows that
\begin{equation}
  T(n,S) = \left(\frac{\pi^2\hbar^3}{\gamma \sqrt{2}}\right)^{2/3}\frac{n^{2/3}}{m^*(n)}\frac{1}{F_{1/2}^{2/3}[f(S)]}
\end{equation}
and
\begin{equation}
  \left.\frac{\partial T(n,S)}{\partial n}\right|_S = \frac{2Q(n)}{3n}T(n,S) 
\end{equation}
with $Q(n)$ given by Eq. (\ref{Q}). The total energy density is given by
\begin{equation}
  \epsilon(n,S) = \mathcal{H}(n,S) + mn
\end{equation}
and thus
\begin{equation}
  \left.\frac{\partial \epsilon}{\partial n}\right|_S = \frac{5}{3n}\frac{\hbar^2}{2}\frac{\tau(n,S)}{m^*(n)}
                                                       \left[1-\frac{3n}{5m^*}\frac{dm^*}{dn}\right] + \frac{dV}{dn} + m .
\end{equation}
The pressure is obtained from
\begin{eqnarray}
  P(n,S) &=& n \left.\frac{\partial \epsilon}{\partial n}\right|_S - \epsilon(n,S)  \\
         &=& \frac{2}{3}\frac{\hbar^2}{2}\frac{Q(n)}{m^*(n)}\tau(n,S) + n \frac{dV}{dn} - V(n)
\end{eqnarray}
and therefore
\begin{eqnarray}
  \left.\frac{\partial P}{\partial n}\right|_S &=&  n \frac{d^2V}{dn^2}+ \frac{10}{9n}\frac{\hbar^2}{2}\frac{Q(n)}{m^*(n)}\tau(n,S) \nonumber  \\
                                               &\times& \left[1-\frac{3n}{5m^*(n)}\frac{dm^*}{dn} + \frac{3n}{5Q(n)}\frac{dQ}{dn}\right]  .
\end{eqnarray}

In the causality-fixing regime $[n\ge n_f(S)]$, Eqs. (\ref{veps}) and (\ref{pp}) imply
\begin{eqnarray}
  \left.\frac{\partial \epsilon}{\partial n}\right|_S &=& \left.\frac{\partial \varepsilon}{\partial n}\right|_S + m
                                     = \frac{(\varepsilon_f+mn_f+P_f)}{n_f}\left(\frac{n}{n_f}\right)^{\beta_f}  \\
  \left.\frac{\partial P}{\partial n}\right|_S &=& \frac{\beta_f}{n_f}(\varepsilon_f+mn_f+P_f)\left(\frac{n}{n_f}\right)^{\beta_f}  .
\end{eqnarray}
Correspondingly, $(c_s/c)^2=\beta_f$ as indicated earlier.

For the conversion of $\epsilon$, $P$, and $c_s$ to the $(n,T)$ variables, we must first express the entropy in terms of $n$ and $T$. This is
accomplished by solving
\begin{equation}
  n = \frac{\gamma}{2\pi^2}\left[\frac{2m^*(n)T}{\hbar^2}\right]^{3/2}F_{1/2}(f)
\end{equation}
for $f(n,T)$ which is then used as input in the functions that appear in Eq. (\ref{sjel}).

\section{Results}
\label{Results}

\begin{figure*}[htb]
\centering
\makebox[0pt][c]{
\hspace{-1.5cm}
\begin{minipage}[b]{0.33\linewidth}
\centering
\includegraphics[width=5cm]{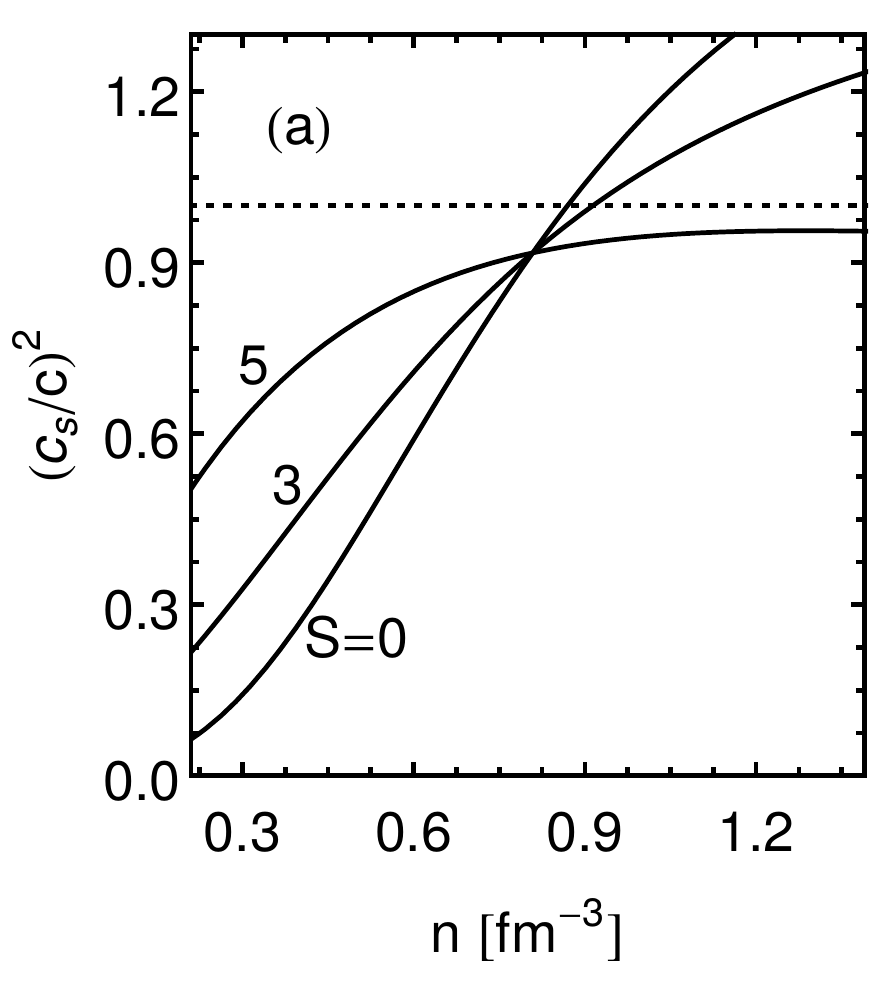}
\end{minipage}
\hspace{-0.5cm}
\begin{minipage}[b]{0.33\linewidth}
\centering
\includegraphics[width=5cm]{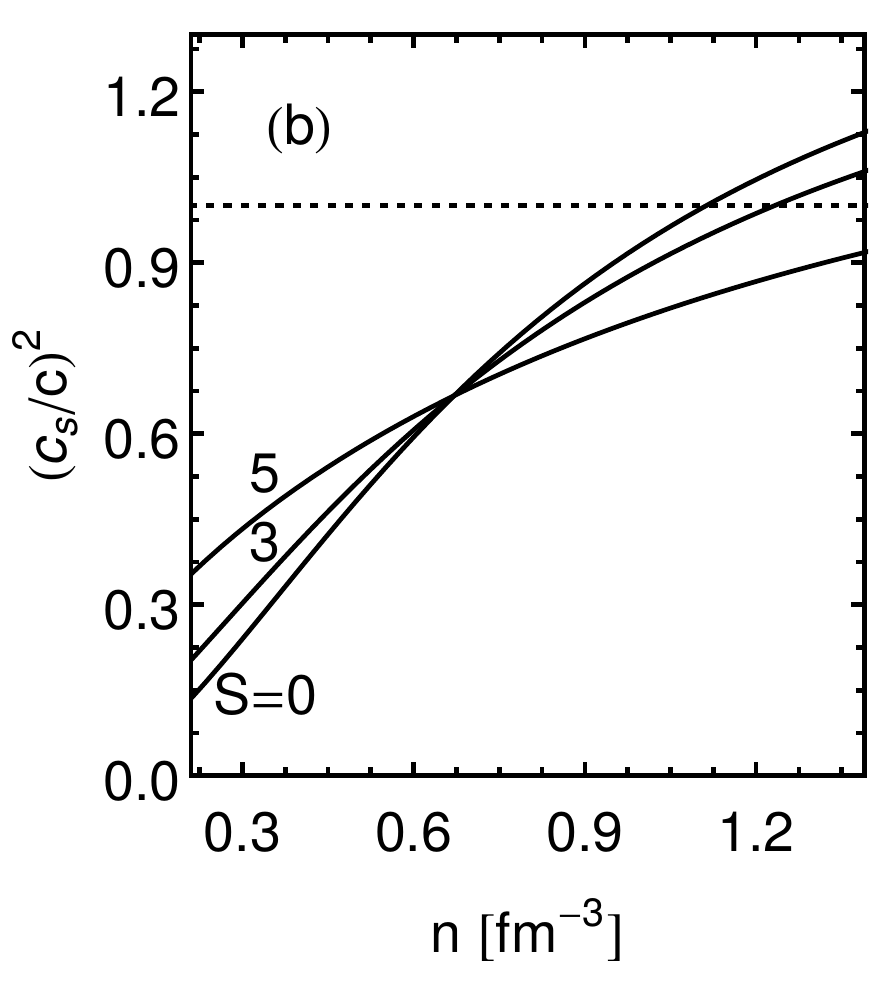}
\end{minipage}
\hspace{-0.5cm}
\begin{minipage}[b]{0.32\linewidth}
\centering
\includegraphics[width=5cm]{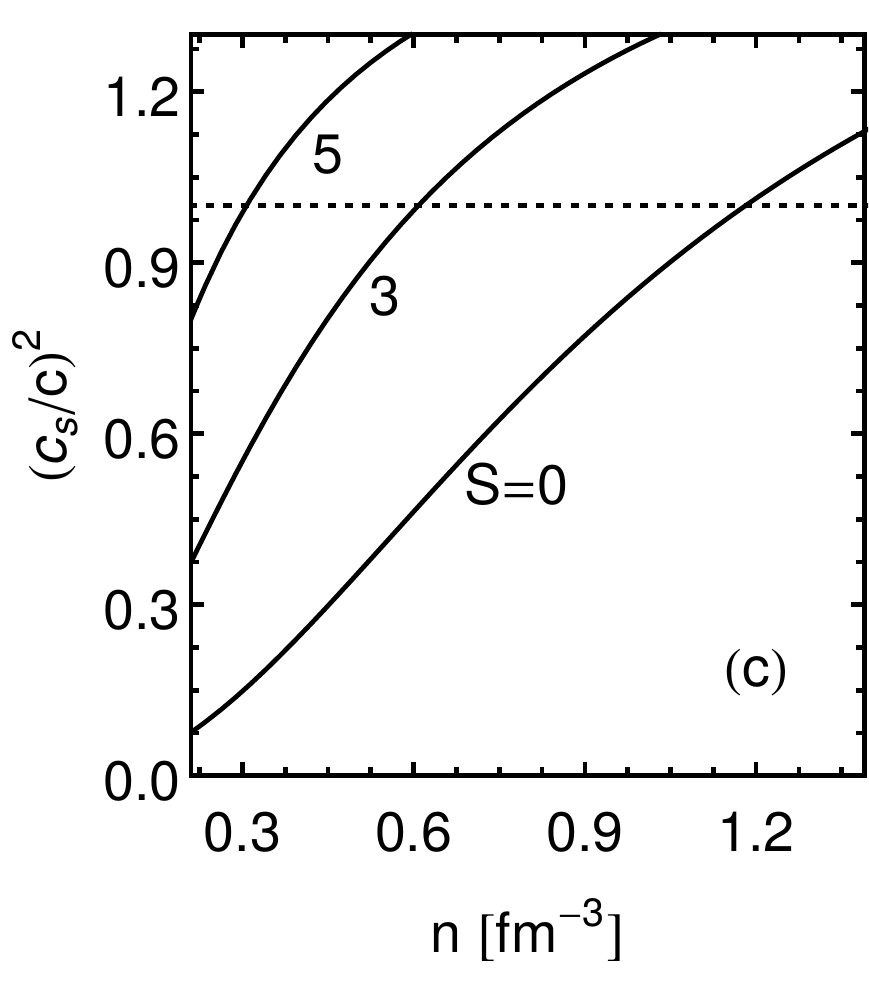}
\end{minipage}}
\vskip -0.25cm
\caption{Squared speed of sound vs density in PNM for the models of APR (a), LS (b), and SLy4 (c) at fixed entropy.}
\label{cs_acausal}
\end{figure*}

In this section, we present results pertaining to the speed of sound for the PNM models of APR \cite{APR,APRus}, LS \cite{LS}, and SLy4 \cite{sly4} and the
alterations our causality-enforcing scheme causes to the properties of the neutron stars in their maximum-mass  configurations. This configuration reaches the
largest central density and therefore it is the setting where the effects of  causality implementation will be most apparent. Our results also illustrate how
various thermodynamic functions are modified by this approach in the case of PNM for the EOS of APR.

Figure \ref{cs_acausal} shows the squared speed of sound of the three models for PNM for different values of the entropy. Results for APR and LS
are qualitatively similar in that for densities lower than a certain density $n_X$, the higher-entropy curves lie higher whereas the situation is
reversed for densities $n>n_X$. Consequently, causality for finite entropies (temperatures) is violated at  densities that are higher than those at zero
temperature for these two models. The intersection point at intermediate densities is common to all curves (for each model)
and thus independent of the entropy. Its value is obtained by solving
\begin{equation}
  \left(\frac{c_s}{c}\right)^2_{S=0} - \left(\frac{c_s}{c}\right)^2_{S\gg1} = 0 \,,
\end{equation}
where the first term refers to the squared speed of sound in cold matter and the second term is given by Eq. (\ref{csindept}).

The speed of sound of SLy4 on the other hand, is a monotonically increasing function of the entropy and hence the
causality-violating density $n_a$ decreases with increasing entropy. The $n_a$ for the three models at $S$=0, 3, and 5
as well as the fixed points $n_X$ of APR and LS are given in Table \ref{nanx}.

\begin{table}[!h]
\begin{ruledtabular} 
\newcolumntype{a}{D{.}{.}{3,13}}
\begin{tabular}{lccc}    
Property & APR & LS & SLy4 \\
\hline
$n_{a,0}$(fm$^{-3}$)        & 0.870(0.841)       &   1.112(1.092)        &    1.181(1.298)                       \\
$n_{a,3}$(fm$^{-3}$)        & 0.914(0.849)       &   1.232(1.165)        &    0.608(0.814)                       \\
$n_{a,5}$(fm$^{-3}$)        & 2.710(0.994)       &   1.774(1.478)        &    0.307(0.454)                       \\
$n_X$(fm$^{-3}$)           & 0.809(0.830)       &   0.671(0.708)        &    N/A                       \\
\end{tabular}
\caption{Densities at which causality is violated at S=0, 3, and 5 for APR, LS, and SLy4 in their PNM(SNM) configuration
  and intersection density (where applicable).}
\label{nanx}
\end{ruledtabular}
\end{table}

\begin{figure}[htb]
\centerline{\includegraphics[width=9.2cm]{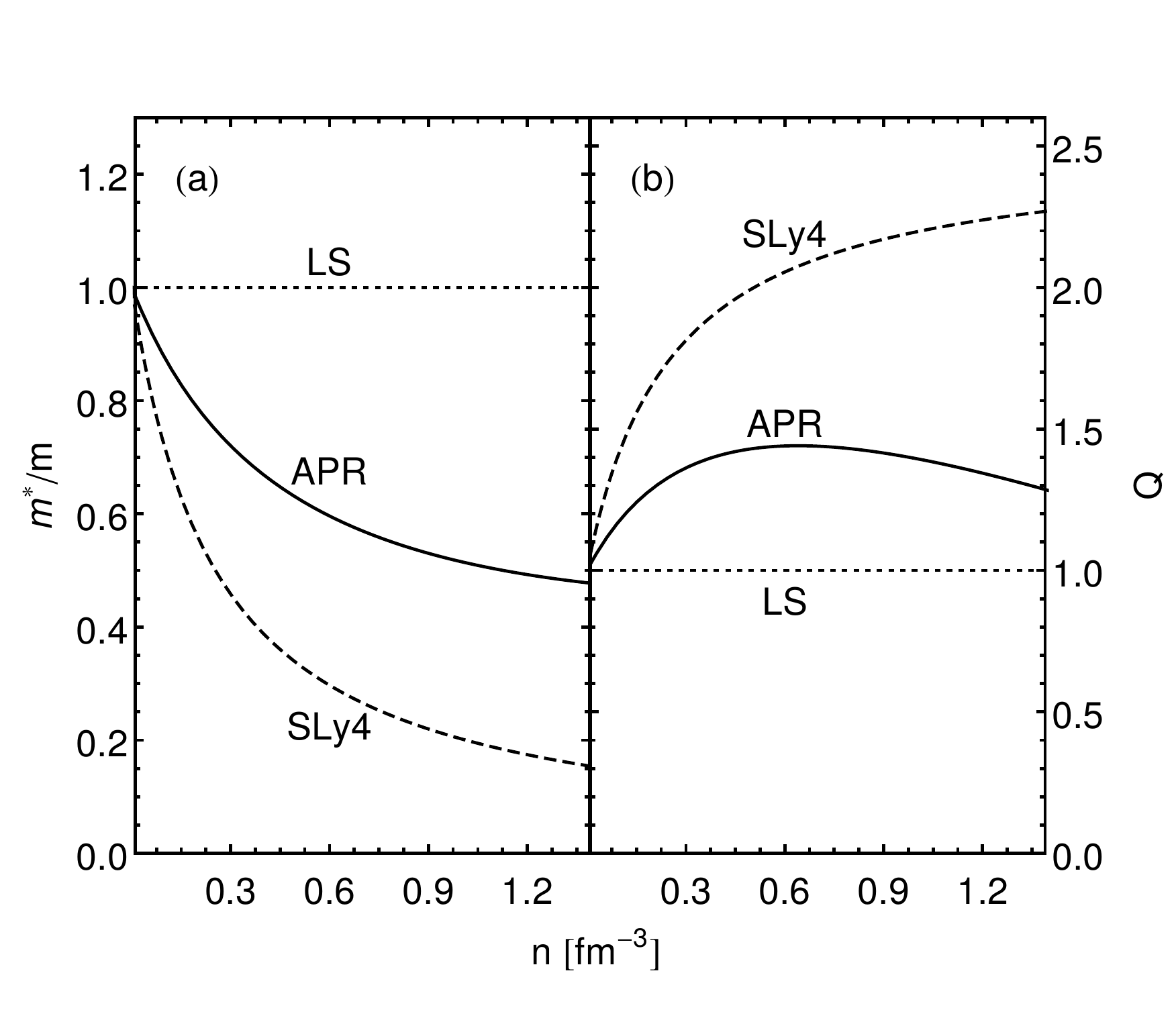}}
\vspace*{-0.25in}
\caption{The effective mass ratio $m^*/m$ and the quantity $Q=1-(3n/2m^*)dn/dm^*$ vs density  in PNM for the APR, LS and SLy4 models.}
\label{msQ}
\end{figure}

The differences in the results of $c_s^2$ for the three models are related to the behaviors of the effective masses and their 
derivatives with respect to density as reflected in the function $Q(n)$ and its derivative with respect to density. Figure \ref{msQ} shows
results of $m^*/m$ and $Q(n)$ vs density. For the LS model  here, $Q(n)=1$ as $m^*(n)=m$, the vacuum nucleon mass. For the APR and SLy4 models, $m^*/m$
decreases monotonically with density, the variation in the latter case being substantially more than for the former.  These variations are in turn
reflected in the behaviors of $Q(n)$ with $n$ for these two models. These results clearly indicate the crucial role of the density dependence of the
effective mass on the speed of sound in hot, dense matter.

Tables \ref{aprns}-\ref{slyns} list the maximum mass and the corresponding radius and central number density $n_c$,
total energy density $\epsilon_c$, and pressure $P_c$ for different values of $\beta_f$ at $T=0$. The last column in each table displays
these quantities as obtained using the original causality-violating EOS. As $\beta_f$ is increased toward 1, $M_{max}$, $n_c$, $\epsilon_c$, and $P_c$
approach their pre-implementation values from below whereas $R_{max}$ does so from above. Changes to these quantities (compared to
pre-implementation) are small; about 10$\%$ even for $\beta_f=0.5$, with the notable exception of the central pressure $P_c$ which nearly halves. For LS,
the $\beta_f=0.9$ star is identical to the original because $n_f(\beta_f=0.9)$ exceeds the central density of the star. A similar
consideration applies for SLy4 for which $n_f(\beta_f=0.9)$ is relatively close to $n_c$.

\begin{table}[!h]
\begin{ruledtabular}
\newcolumntype{a}{D{.}{.}{-1}}
\begin{tabular}{laaaa}
\backslashbox{Prop.}{$\beta_f$} & \multicolumn{1}{c}{0.5} &\multicolumn{1}{c}{0.7} &\multicolumn{1}{c}{0.9}& \multicolumn{1}{c}{Not Fixed} \\
\hline
$n_f $(fm$^{-3}$)           & 0.547 & 0.667 & 0.796 & N/A\\
$M_{max} (M_{\odot})$        & 2.00 & 2.13 & 2.18 & 2.20 \\
$R_{max} $(km)              & 10.61 & 10.46 & 10.31 & 10.16 \\
$n_c $(fm$^{-3}$)           & 1.107 & 1.096 & 1.101 & 1.111 \\
$\epsilon_c$(MeV fm$^{-3}$) & 1398.6 & 1433.3 & 1472.0 & 1507.0 \\
$P_c $(MeV fm$^{-3}$)       & 516.0 & 691.9 & 851.1 & 1005.1 \\
\end{tabular}
\caption{PNM neutron star properties for the APR model for different values of $\beta_f$ at $T=0$.}
\label{aprns}
\end{ruledtabular}
\end{table}

\begin{table}[!h]
\begin{ruledtabular}
\newcolumntype{a}{D{.}{.}{-1}}
\begin{tabular}{laaaa}
\backslashbox{Prop.}{$\beta_f$} & \multicolumn{1}{c}{0.5} &\multicolumn{1}{c}{0.7} &\multicolumn{1}{c}{0.9}& \multicolumn{1}{c}{Not Fixed} \\
\hline
$n_f $(fm$^{-3}$)           & 0.515 & 0.705 & 0.951 & N/A\\
$M_{max} (M_{\odot})$        & 2.23 & 2.29 & 2.30 & 2.30 \\
$R_{max} $(km)              & 12.03 & 11.70 & 11.58 & 11.58 \\
$n_c $(fm$^{-3}$)           & 0.875 & 0.906 & 0.915 & 0.915 \\
$\epsilon_c$(MeV fm$^{-3}$) & 1101.6 & 1175.9 & 1197.7 & 1197.7 \\
$P_c $(MeV fm$^{-3}$)       & 398.0 & 534.0 & 584.1 & 584.1 \\
\end{tabular}
\caption{Same as Table \ref{aprns}, but for the LS model.}
\label{lsns}
\end{ruledtabular}
\end{table}

\begin{table}[!h]
\begin{ruledtabular}
\newcolumntype{a}{D{.}{.}{-1}}
\begin{tabular}{laaaa}
\backslashbox{Prop.}{$\beta_f$} & \multicolumn{1}{c}{0.5} &\multicolumn{1}{c}{0.7} &\multicolumn{1}{c}{0.9}& \multicolumn{1}{c}{Not Fixed} \\
\hline
$n_f $(fm$^{-3}$)           & 0.634 & 0.825 & 1.048 & N/A\\
$M_{max} (M_{\odot})$        & 1.95 & 2.03 & 2.05 & 2.05 \\
$R_{max} $(km)              & 10.46 & 10.20 & 10.03 & 10.00 \\
$n_c $(fm$^{-3}$)           & 1.150 & 1.173 & 1.195 & 1.196 \\
$\epsilon_c$(MeV fm$^{-3}$) & 1454.2 & 1534.4 & 1590.7 & 1594.0 \\
$P_c $(MeV fm$^{-3}$)       & 530.0 & 717.1 & 852.6 & 872.1 \\
\end{tabular}
\caption{Same as Table \ref{aprns}, but for the SLy4 model.}
\label{slyns}
\end{ruledtabular}
\end{table}

\begin{figure}[htb]
\centerline{\includegraphics[width=9.2cm]{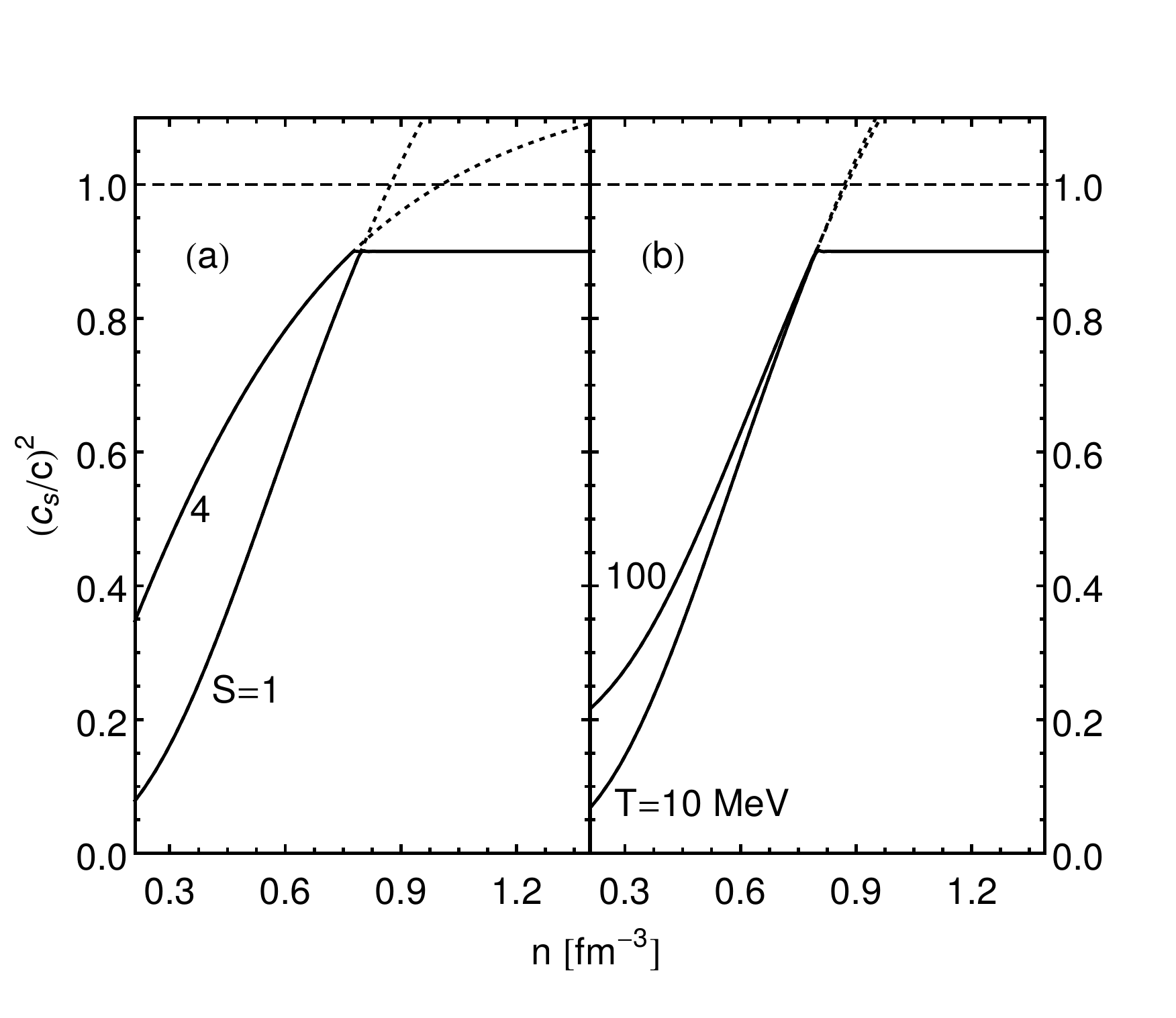}}
\vspace*{-0.25in}
\caption{Squared speed of sound in PNM for the APR model with (solid curves) and without (dotted curves) causality enforced with $\beta_F=0.9$ for fixed entropy (a) and
  temperature (b)  vs density.}
\label{apr_cs}
\end{figure}

Figure \ref{apr_cs} shows how the squared speed of sound of APR (PNM) is altered by our method as a function of the density for fixed entropy (left panel)
and for fixed temperature (right panel) for a fixed $\beta_f=0.9$. That the different curves appear to be causally fixed at the same density is a consequence of the (accidental)
fact that, for APR, $n_f(\beta_f=0.9)\simeq n_X$. As a caution we point out that the problematic implementation of \cite{APRus} will \textit{appear} correct if
one chooses $\beta_f = \beta_f(n_X)$.

\begin{figure}[htb]
\centerline{\includegraphics[width=9.2cm]{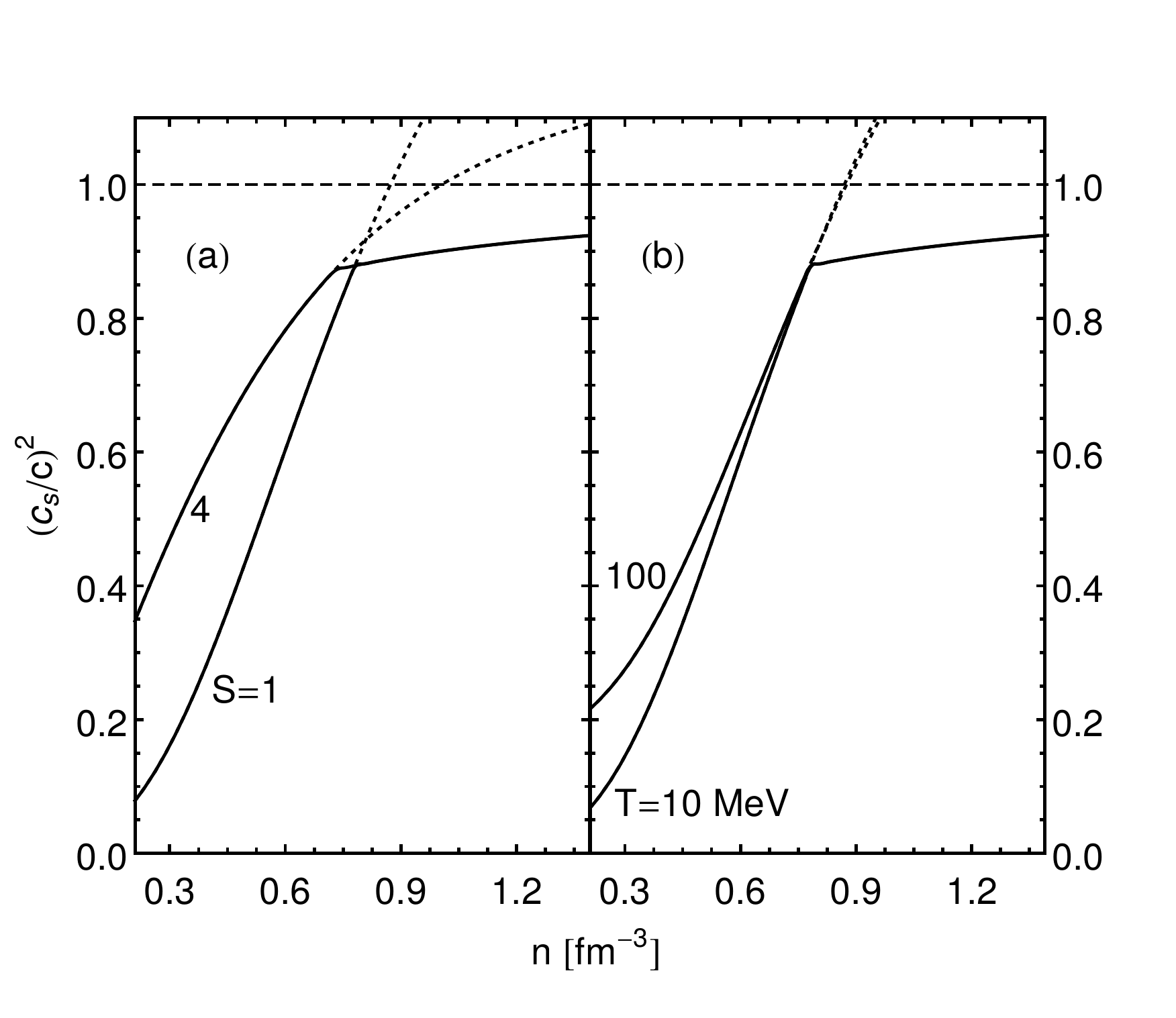}}
\vspace*{-0.25in}
\caption{
Same as Fig. \ref{apr_cs}, but with the density-dependent  $\beta_f$ given by Eq. (\ref{beta2}).}
\label{apr_cs2}  
\end{figure}

A notable feature of  the results in Fig. \ref{apr_cs} is that $(c_s/c)^2=0.9$ for all $n\geq n_a$, the density at which acausality sets in for the APR model.
With a density-dependent $\beta_f$, a gradual approach of $(c_s/c)^2$ to 1 may be achieved. Figure \ref{apr_cs2} shows results with the density-dependent $\beta_f$
given by Eq. (\ref{beta2}). As noted  earlier, many other possibilities also exit as long as one can find a tractable, preferably analytical, solution to Eq. ~(\ref{1DE}). 

\begin{figure}[htb]
\centerline{\includegraphics[width=9.2cm]{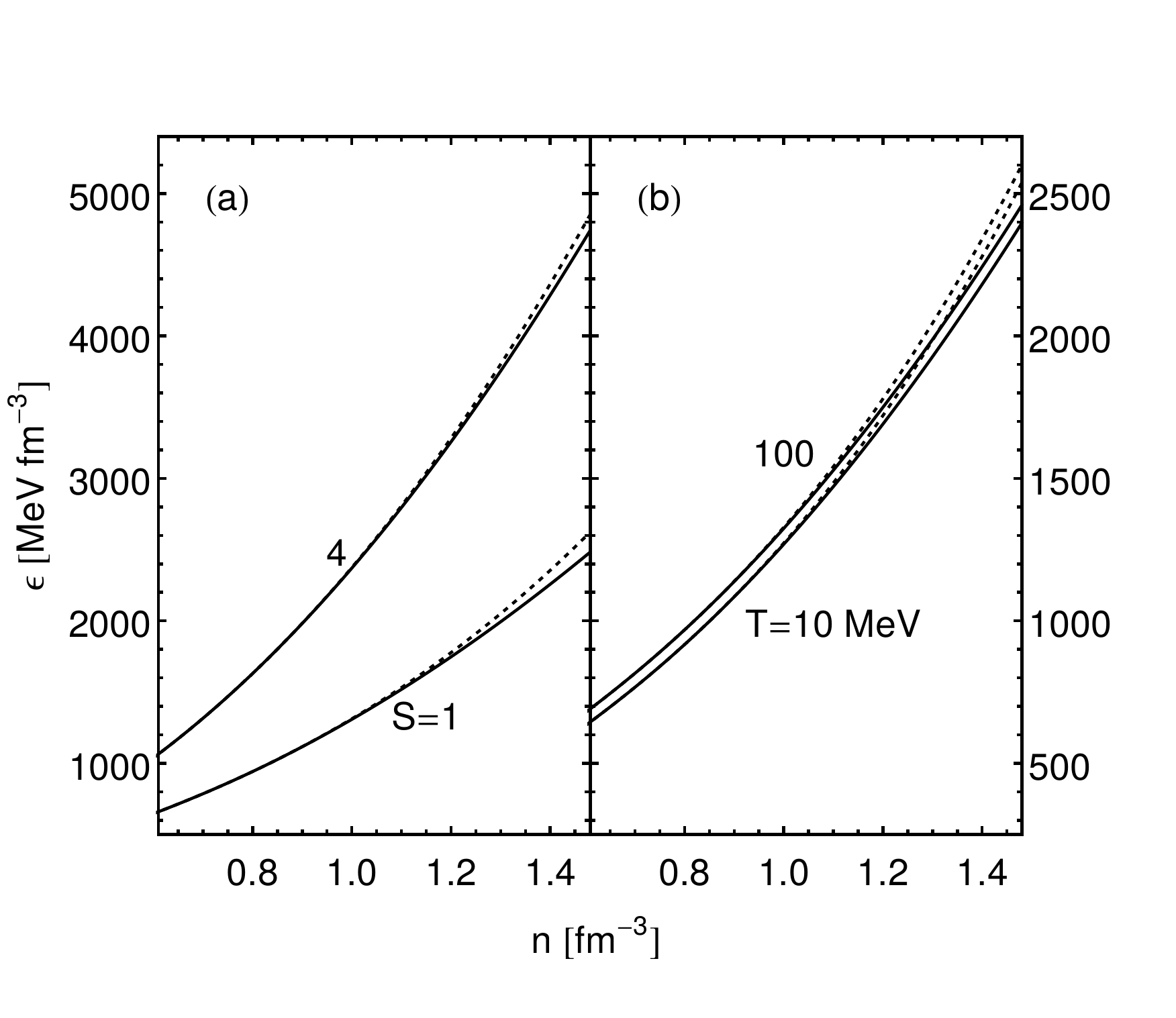}}
\vspace*{-0.25in}
\caption{Total energy density of PNM for the APR model with (solid curves) and without (dotted curves) causality enforced with 
$\beta_f=0.9$ for fixed entropy (a) and temperature (b) vs density.}
\label{apr_eden}
\end{figure}

\begin{figure}
\centerline{\includegraphics[width=9.2cm]{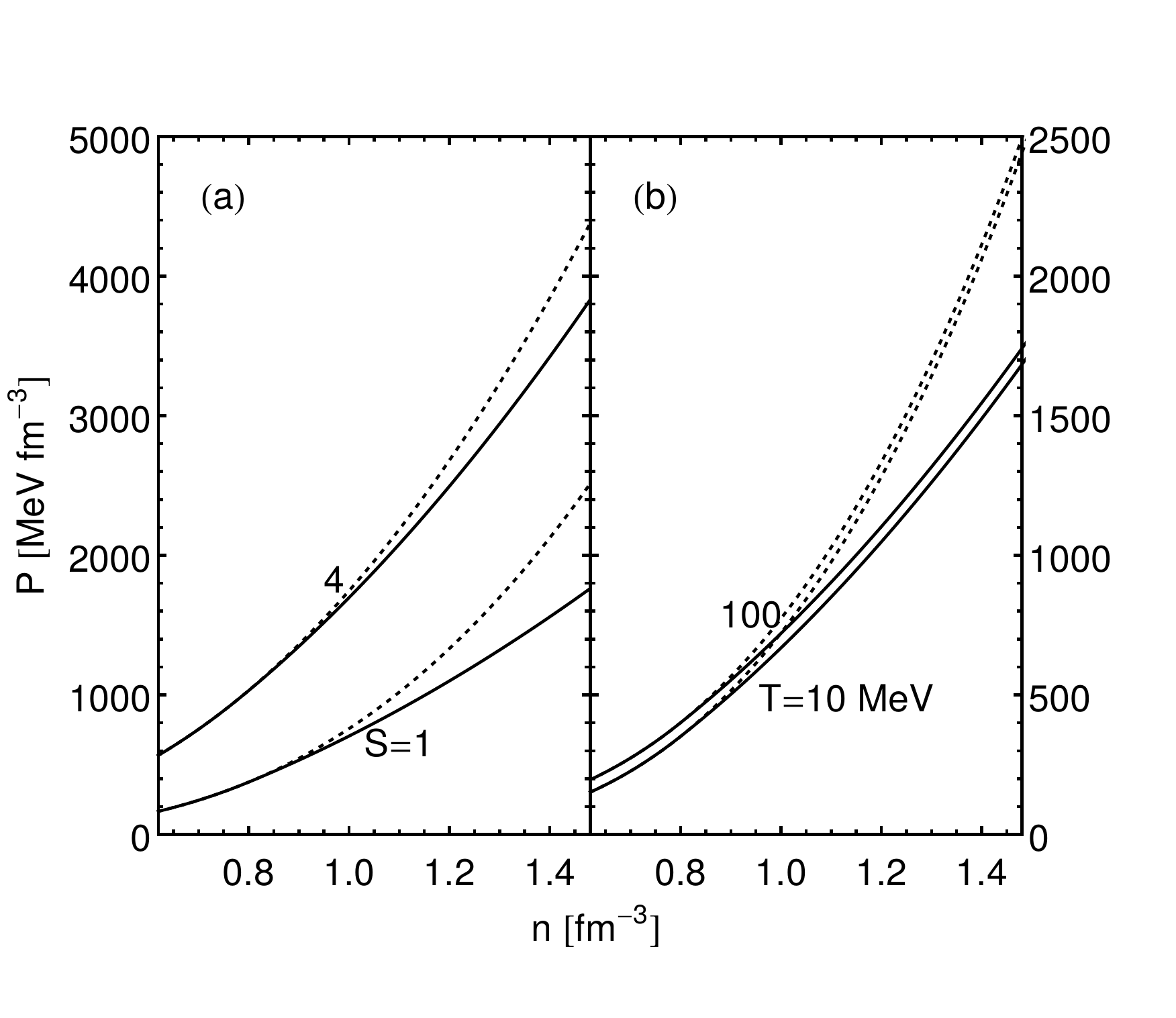}}
\vspace*{-0.25in}
\caption{Pressure of PNM for the APR model with (solid curves) and without (dotted curves) causality enforced with 
$\beta_f=0.9$ for fixed entropy (a) and temperature (b) vs density.} 
\label{apr_pres}
\end{figure}

\begin{figure}[htb]
\centerline{\includegraphics[width=9.2cm]{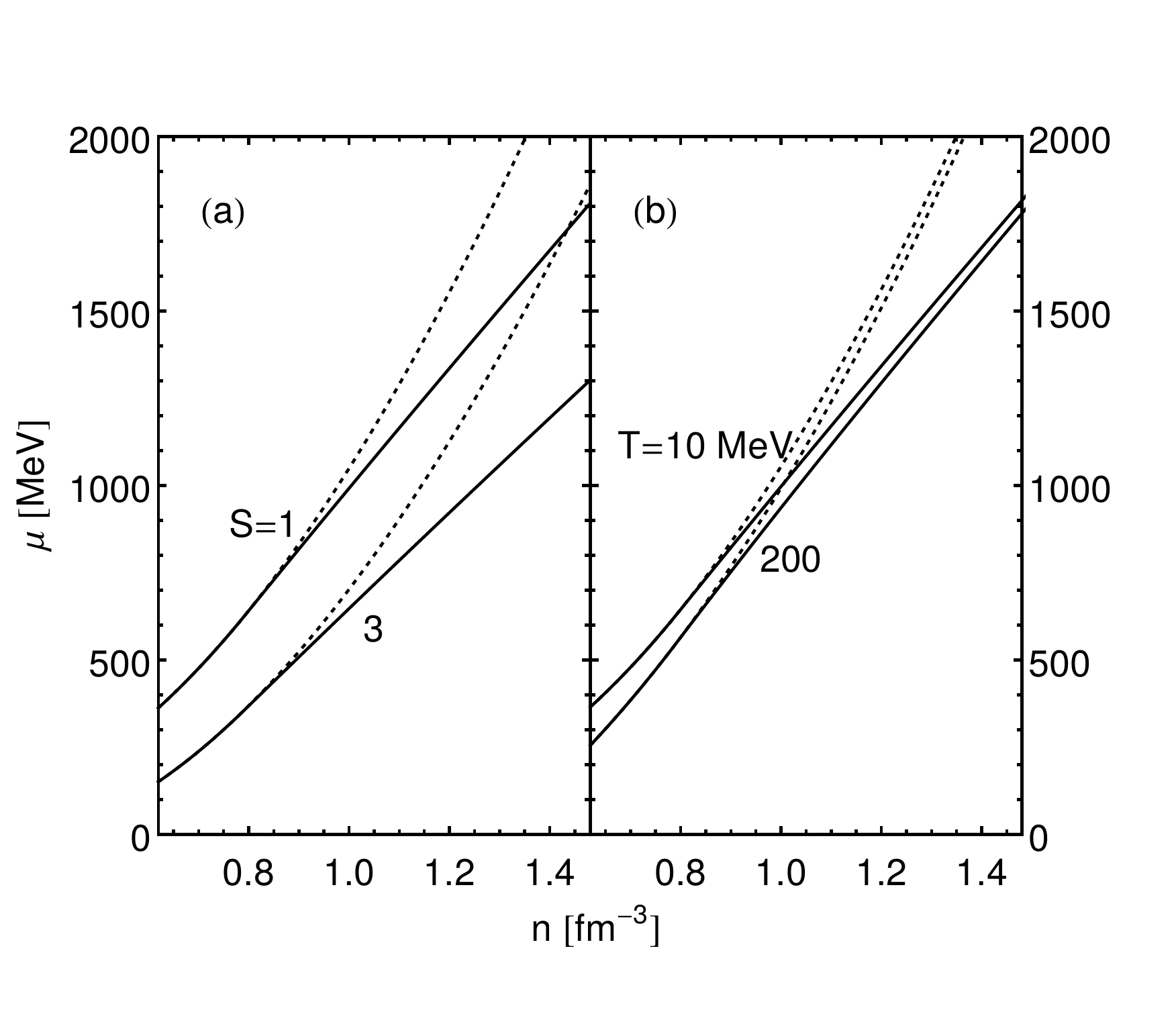}}
\vspace*{-0.25in}
\caption{Chemical potential in PNM for the APR model with (solid curves) and without (dotted curves) causality enforced with 
$\beta_f=0.9$ for fixed entropy (a) and temperature (b) vs density.}
\label{apr_mu}
\end{figure}

\begin{figure}[htb]
\centerline{\includegraphics[width=9.2cm]{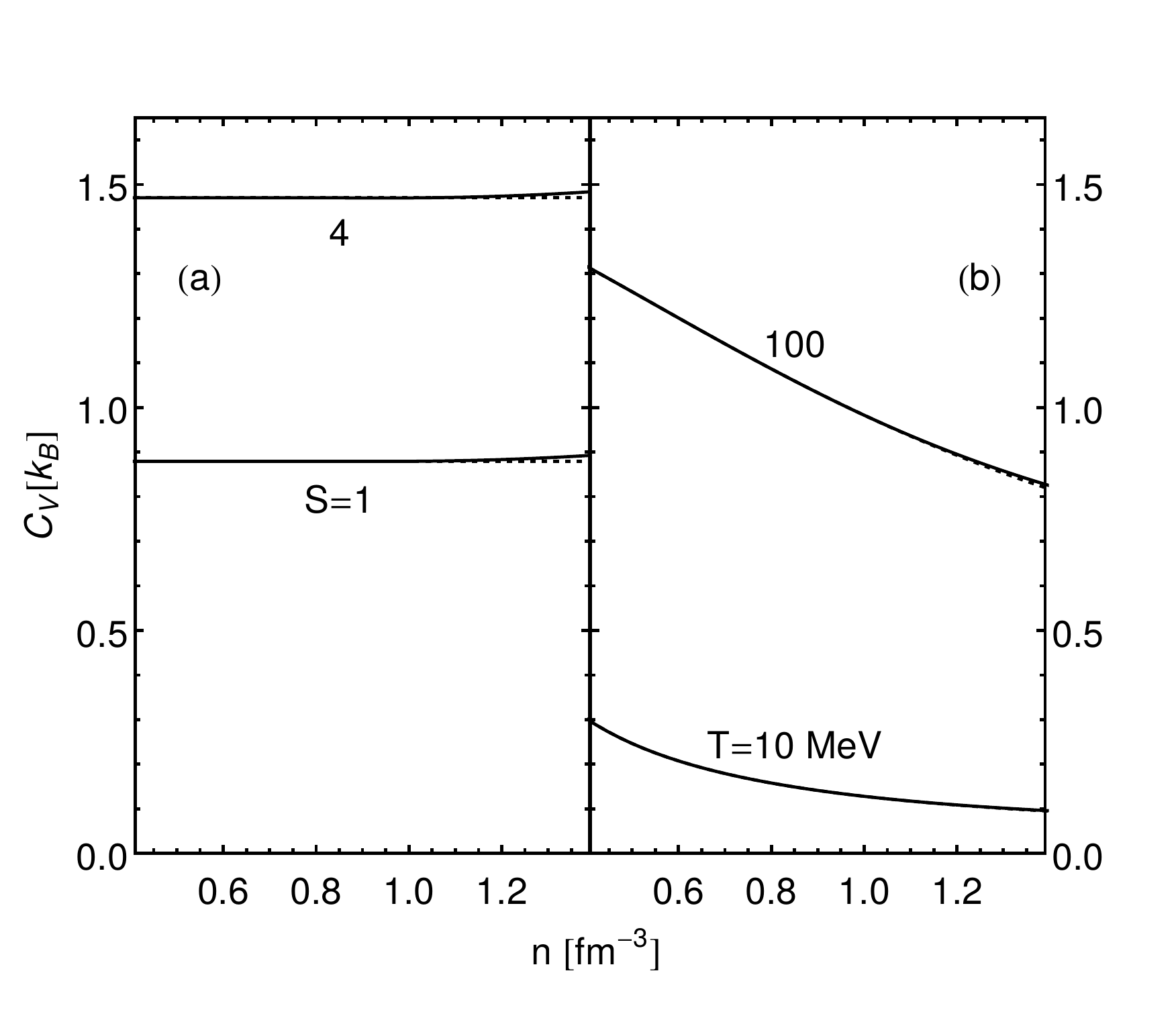}}
\vspace*{-0.25in}
\caption{Specific heat at constant volume in PNM for the APR model with (solid curves) and without (dotted curves) causality enforced with 
$\beta_f=0.9$ for fixed entropy (a) and temperature (b) vs density.}
\label{apr_cv}
\end{figure}

\begin{figure}
\centerline{\includegraphics[width=9.2cm]{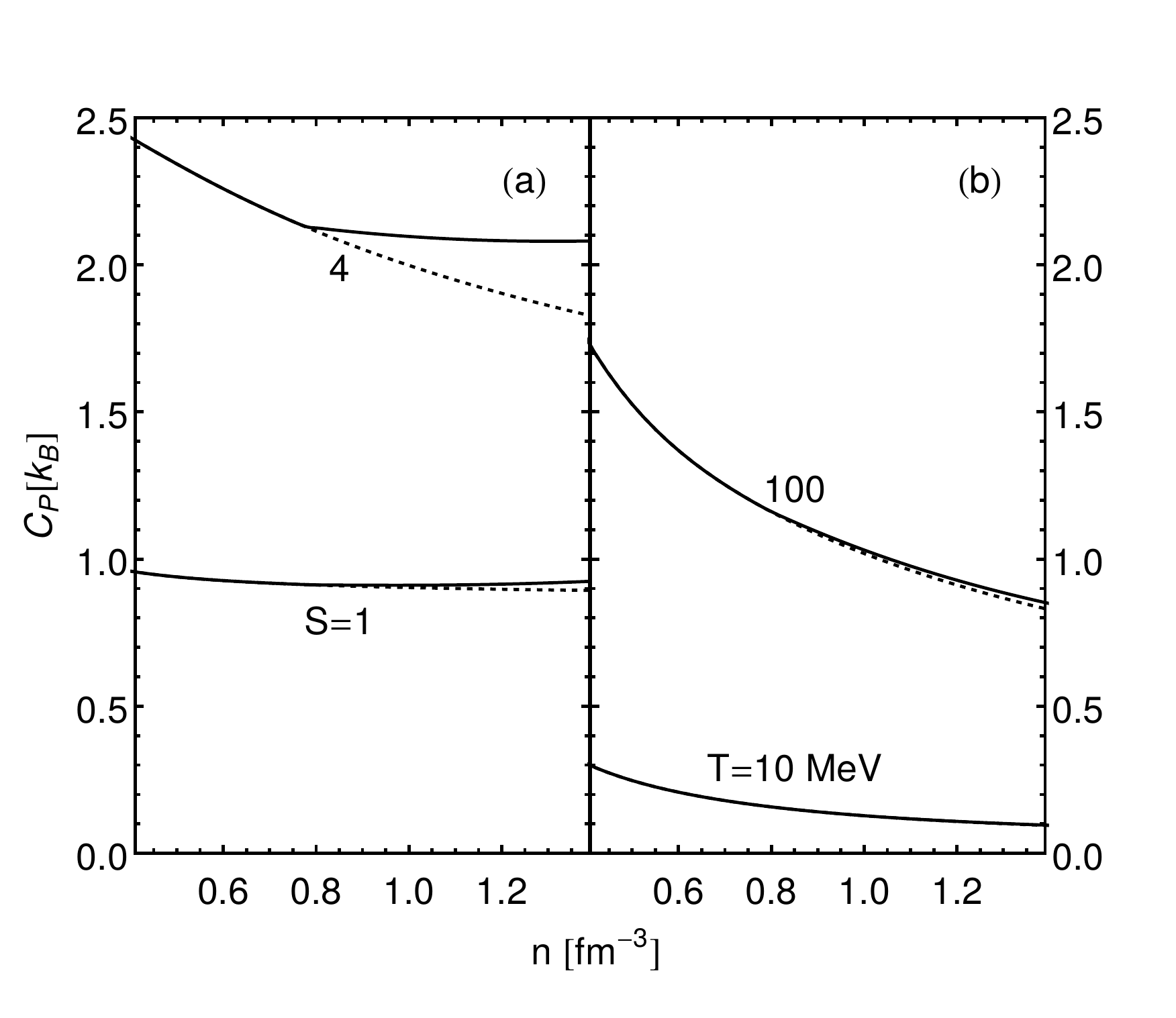}}
\vspace*{-0.25in}
\caption{Specific heat at constant pressure in PNM for the APR model with (solid curves) and without (dotted curves) causality enforced with 
$\beta_f=0.9$ for fixed entropy (a) and temperature (b) vs density.}
\label{apr_cp}
\end{figure}

\begin{figure}[htb]
\centerline{\includegraphics[width=9.2cm]{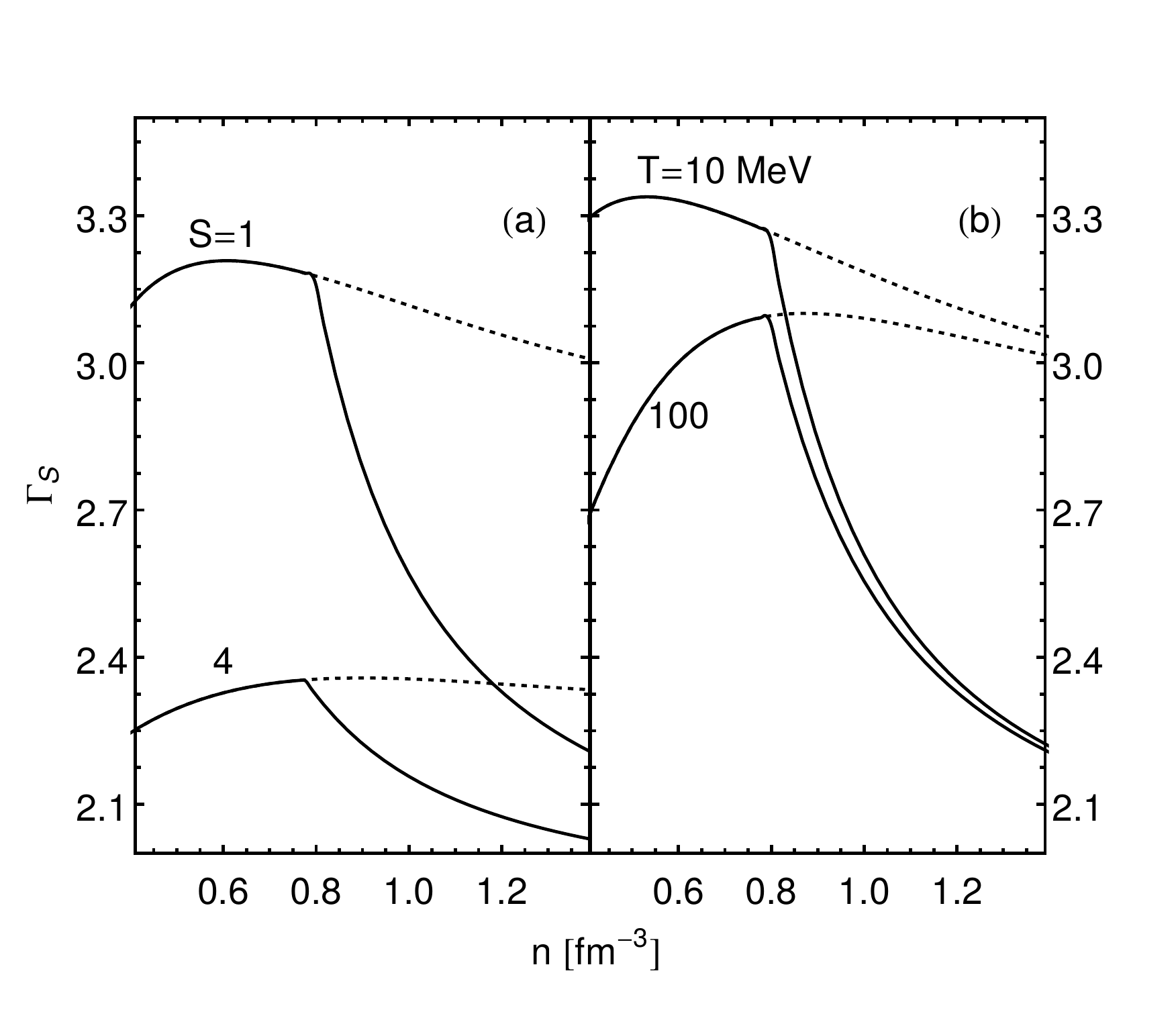}}
\vspace*{-0.25in}
\caption{Adiabatic index in PNM for the APR model with (solid curves) and without (dotted curves) causality enforced with 
$\beta_f=0.9$ for fixed entropy (a) and temperature (b) vs density.}
\label{apr_gs}
\end{figure}

The implementation of causality introduces modifications to the total energy density $\epsilon$, pressure $P$, chemical potential $\mu$, the specific heats $C_V$ and $C_P$, and 
the adiabatic index $\Gamma_S$ which are exhibited in Figs. \ref{apr_eden}-\ref{apr_gs}, respectively. All of these results correspond to $\beta_f=0.9$. These modifications
occur at high densities and are more pronounced for quantities ($P, ~\mu,~\Gamma_S$ and $C_P$) that involve density derivatives of the energy.  This observation is in
accordance with that made earlier regarding the central pressure of neutron stars. Results corresponding to Eq. (\ref{beta2}) are nearly identical and are not shown here.

\subsection*{Comparison with finite-range force models}

\begin{figure}[htb]
\centerline{\includegraphics[width=9.2cm]{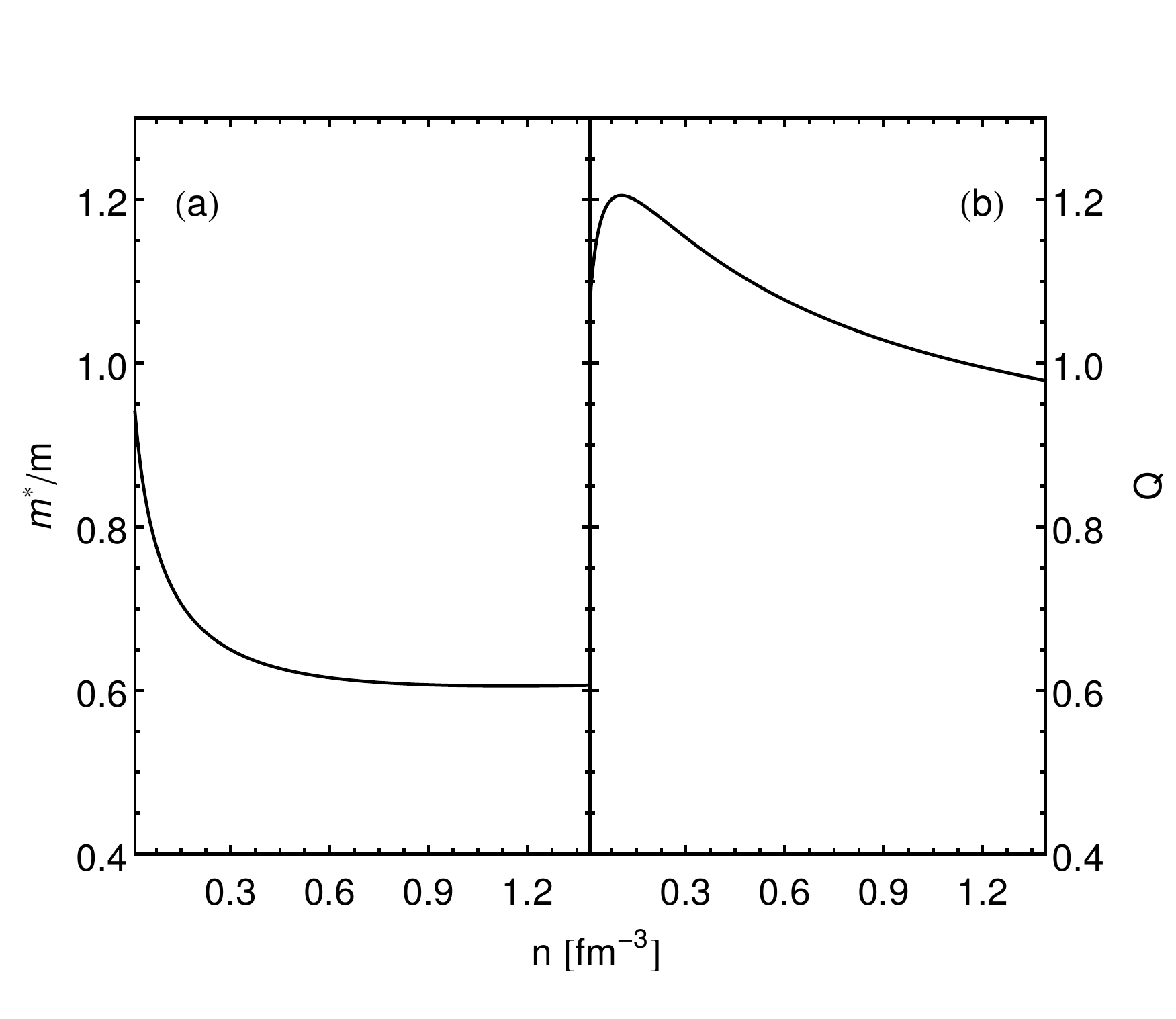}}
\vspace*{-0.25in}
\caption{The effective mass ratio $m^*/m$ and the quantity $Q=1-(3n/2m^*)dn/dm^*$ vs density in PNM for the BPAL33 model.}
\label{bpal_msq}
\end{figure}

Here we contrast the above results for $(c_s/c)^2$ with those of a nonrelativistic potential model with finite-range forces at finite temperature  studied in
detail in Ref. \cite{Prakash97}, where results for $(c_s/c)^2$ were, however, not shown. For PNM, the energy density in this model is 
\begin{eqnarray}
\epsilon = 2 &\int& \frac {d^3k}{(2\pi)^3} \frac {\hbar^2 k^2}{2m}~f + Au^2 + \frac {Bu^\sigma}{1+B^\prime u^{\sigma-1}} 
\nonumber \\
&+&   u \sum_{i=1,2} C_i ~2  \int  \frac {d^3k}{(2\pi)^3} \frac {1} {[1+(k/\Lambda_i)^2]}~f \,,
\label{BPAL33}
\end{eqnarray}
where $u=n/n_s$, $f$ is the usual Fermi-Dirac distribution function at finite $T$, and the parameters $A,~B,~\sigma,~C_i,~B^\prime,~{\rm and}~ \Lambda_i$ are determined
from constraints provided by the empirical properties of nuclear matter at $n_s$. Referred to as BPAL33 in Ref. \cite{Prakash97}, their numerical values are:
$A=1.627, ~B=8.908, ~B^{\prime}=0.422, ~C_1=-106.7 ~\mbox{MeV}, ~C_2=6.544 ~\mbox{MeV}, ~\Lambda_1=1.5\hbar k_F^{(0)}, ~\mbox{and} ~\Lambda_2=3.0\hbar k_F^{(0)}$
with $k_F^{(0)}=(3\pi^2n_s/2)^{1/3}$. Note the redefinition of parameters here from those in the original reference. The energy density in Eq. (\ref{BPAL33}) differs
from that of zero-range Skyrme-like models including the APR model in two respects. First, the term encapsulating the influence of higher-than two body forces is
such that it does not lead to an acausal behavior at $T=0$. Secondly, the finite-range terms lead to an effective mass 
\begin{equation}
\frac {m^*}{m} = \left[1+ \sum_{i=1,2} \alpha_i u \left(1+ \frac {(2u)^{2/3} }{R_i^2}  \right) \right]^{-1} \,,
\end{equation}
where $R_i=\Lambda_i/(\hbar k_F^{(0)})$, that saturates for $n >> n_s$ as shown in Fig. \ref{bpal_msq}. As a result, $Q<1.2$ and $dQ/dn<0$ for $n > n_s$ which implies
that $c_s<c$ in the limit $T\rightarrow \infty$ [Eq. (\ref{csindept})]. This means that $c_s < c$ for all $T$ being that the possible paths that $c_s$ can traverse
in ($n,T$) are bounded by $c_s(T=0)$ and $c_s(T\rightarrow \infty)$. In closing this section we note that, for BPAL33, the intersection density $n_X = 0.853$ fm$^{-3}$.

\begin{figure}
\centerline{\includegraphics[width=9.2cm]{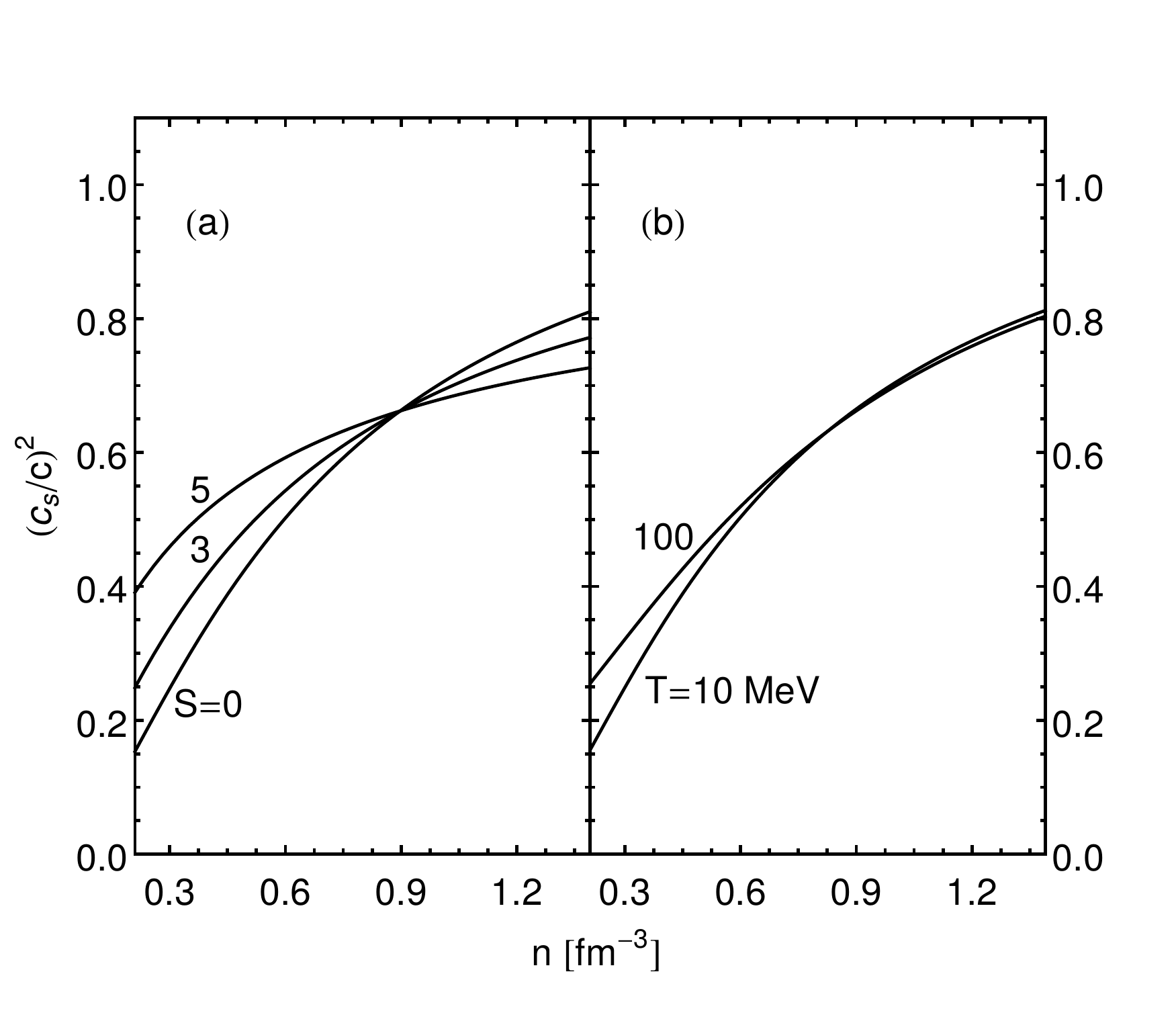}}
\vspace*{-0.25in}
\caption{Squared speed of sound in PNM for the BPAL33 model for fixed entropy (a) and temperature (b) vs density.}
\label{bpal_cs}
\end{figure}

To preserve causality, two lessons, of much value to first principle microscopic calculations hot and dense matter, are learned from the results above.
First,  contributions from higher than two-body forces must be screened at high density at the $T=0$ level. Secondly,  the nucleon effective mass, which
controls thermal effects, must not rapidly decrease with density as in some Skyrme-like models that employ only contact interactions. The use of finite-range
forces (as in first principle calculations of dense matter), which tends to saturate the nucleon effective mass, mitigates the influence of thermal effects
in making EOS's acausal. \\

\section{Summary and Conclusions}
\label{Conclusions}

In this work, we have proposed a method by which nonrelativistic EOS's that become acausal at high densities can be modified so that they remain causal at all densities
and entropies/temperatures. This approach is easily implemented and computationally straightforward; its most important feature is thermodynamic consistency.
Illustrative calculations are presented both for a fixed value of the speed of sound $c_s$ in the ``causality-enforcement'' region as well as for continuous functions
of density and entropy per baryon $(n,S)$ which approach $c$ asymptotically from below.

As examples, we have explored consequences of enforcing causality to the attributes of maximum-mass neutron star configurations in pure neutron matter  for the APR, LS,
and SLy4 models. The EOS functions of the APR model are presented for entropies per baryon of relevance to astrophysical simulations before and after enforcing causality.
Our principal findings are summarized below.
 
Insofar as our choice for the ``new'' speed of sound $c_s$ is close to $c$, we find that both cold and finite-$T$ properties associated with the energy density,
$\varepsilon$, and  the specific heat at constant volume, $C_V$, are relatively weakly affected after enforcing causality. However, properties such as the pressure, $P$,
the chemical potential, $\mu$, and the specific heat at constant pressure, $C_P$, which are related to density derivatives of the energy exhibit larger variations compared
to $\varepsilon$ and $C_V$.  At $T=0$, the basic characteristics of PNM-NS configurations such as their central density, $n_c$, the maximum mass, $M_{max}$, and the radius
of the maximum configuration, $R_{max}$, are not greatly affected by enforcing causality. However, for models (such as SLy4) in which the effective nucleon mass drops
rapidly with density thermal effects cause $c_s$ to exceed $c$ at densities significantly lower than at $T=0$. An interesting finding is that in the extreme nondegenerate
limit, $c_s$ for models with contact interactions such as those considered here decouples from entropy/temperature and is instead determined by the Landau effective mass
and its derivatives with respect to density. 

Finally, our study of a schematic potential model illustrates that in  first principle calculations of hot and dense matter, contributions from higher than two-body
interactions must be screened  and effective masses determined by finite-range forces must saturate at high density to preserve causality.

\section*{Acknowledgments}
Research support for  M.P.  by the U.S. DOE under Grant No. DE-FG02-93ER-40756 is gratefully acknowledged.


\bibliographystyle{h-physrev3}
\bibliography{references}

\end{document}